\documentclass[preprint,3p,latin,12pt,authoryear]{elsarticle}

\usepackage{amssymb}
\usepackage{color}
\usepackage{xcolor}
\usepackage[english]{babel}
\usepackage{amsthm}
\theoremstyle{definition}
\newtheorem{remark}{Remark}
\usepackage{geometry}
\usepackage{siunitx}
\usepackage{hyperref}	 
\hypersetup{colorlinks=true,
	linkcolor=blue,%
	citecolor=blue}

\usepackage{algorithm}
\usepackage{algpseudocode}
\makeatletter  
\newif\if@restonecol  
\makeatother  

\newcommand\tb[1]{\boldsymbol{#1}}
\newcommand\td{\mathrm{d}}
\newcommand\pd{\partial}
\newcommand\sym[1]{\mathrm{sym}\left(#1\right)}
\newcommand\skw[1]{\mathrm{skw}\left(#1\right)}

\usepackage{graphicx}
\graphicspath{{./Figures/}}
\usepackage[normalem]{ulem}

\newcommand{\hl}{\textcolor{black}}
\newcommand\mdl{\bgroup\markoverwith{\textcolor{red}{\rule[1.5ex]{2pt}{0.8pt}}}\ULon}

\usepackage[mathlines]{lineno}
\usepackage{etoolbox} 

\usepackage{caption}
\usepackage{subcaption}
\usepackage{booktabs}
\usepackage{multirow}
\usepackage[flushleft]{threeparttable}
\usepackage{bm}
\usepackage{amsmath}
\usepackage{cases}
\allowdisplaybreaks[4]
\usepackage{enumitem}
\usepackage{lscape}
\usepackage{rotfloat}
\usepackage{pdflscape}
\usepackage{makecell}
\usepackage{orcidlink}

\usepackage{tikz}
\journal{arXiv}

\begin{document}
\sloppy

\begin{frontmatter}

\title{An elasto--viscoplastic thixotropic model for fresh concrete capturing flow--rest transition}
\date{\today}

    \author[1,3]{Jidu Yu}
	\author[2]{Bodhinanda Chandra\corref{mycorrespondingauthor}}\ead{chandra@nus.edu.sg}
    \author[4]{Christopher Wilkes}
	\author[3]{Jidong Zhao}
	\author[1]{Kenichi Soga}

	\cortext[mycorrespondingauthor]{Corresponding author}
	\address[1]{Department of Civil and Environmental Engineering, University of California, Berkeley, CA, 94720, United States}
	\address[2]{Department of Civil and Environmental Engineering, National University of Singapore, Singapore 117576}
	\address[3]{Department of Civil and Environmental Engineering, Hong Kong University of Science and Technology, Kowloon, Hong Kong}
    \address[4]{Ove Arup \& Partners, London, United Kingdom}

\begin{abstract}
The flow properties of fresh concrete are critical in the construction industry, as they directly affect casting quality and the durability of the final structure. Although non-Newtonian fluid models, such as the Bingham model, are widely used to model these flow properties, they often fail to capture key phenomena, including flow stoppage, and frequently rely on ad hoc, non-physical regularization or stabilization techniques to mitigate numerical instabilities at low shear rates. To address these limitations, this study proposes an elasto-viscoplastic constitutive model within the continuum mechanics framework, which treats fresh concrete as a solid-like material with a rate-dependent yield stress. The model inherently captures the transition from elastic response to viscous flow following Bingham rheology, and vice versa, enabling accurate prediction of flow cessation without ad hoc criteria. Additionally, a thixotropy evolution law is incorporated to account for the time-dependent behavior resulting from physical flocculation and shear-induced deflocculation. The proposed model is implemented within the Material Point Method (MPM), whose Lagrangian formulation facilitates tracking of history-dependent variables and robust simulation of large deformation flows. Numerical examples demonstrate the model's effectiveness in reproducing a range of typical concrete flow scenarios, offering a more physically consistent numerical tool for optimizing concrete construction processes and minimizing defects.
\end{abstract}


\end{frontmatter}

\nolinenumbers
\section{Introduction}

\begin{figure}[h!]
    \centering
    \includegraphics[width=1\linewidth]{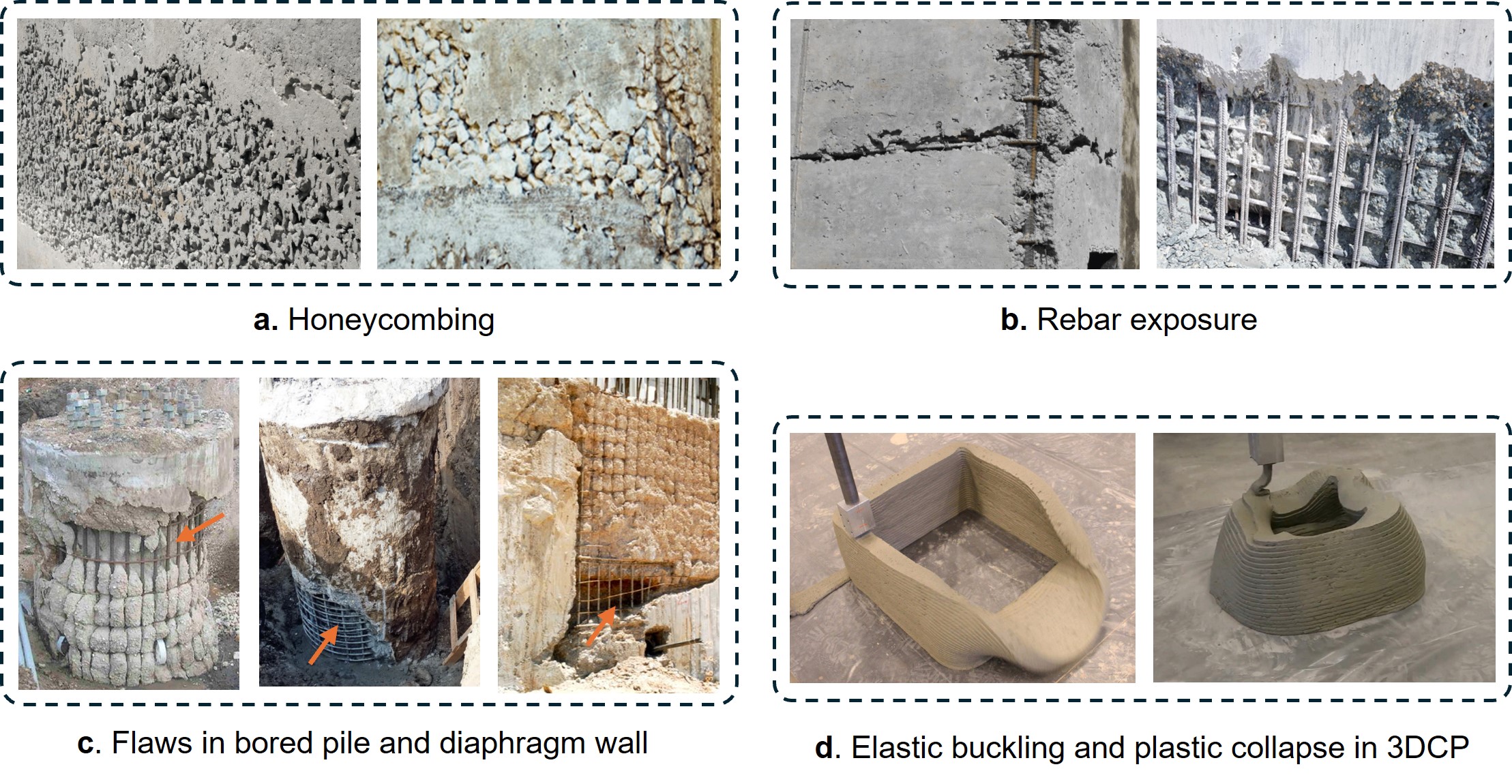}
    \caption{Typical concrete defects and flaws caused by improper casting. In Figure (d), 3DCP refers to 3D Concrete Printing. Figures (a)-(c) are obtained from \cite{civilengineerdk2023, civilengineermag2025, gharpedia2025, elop2023, martinello2018, amir2019}; Figure (d) is adapted from \cite{suiker2020elastic}.}
    \label{fig: concrete_defects}
\end{figure}

Concrete is the most widely used construction material, which serves as a fundamental component of infrastructure development \citep{li2022advanced, environment2018eco}. The flow behavior of fresh concrete, including its workability and flowability, critically influences the quality of placement, compaction, and finishing operations, thereby directly affecting the durability and overall performance of the structure \citep{roussel2007, wallevik2011rheology}. Improper flow during casting can lead to severe defects such as honeycombing, rebar exposure, as well as flaws in the concrete structure, which compromise the long-term structural integrity of the final structural element \citep{sun2021thermal, megid2018effect, suiker2020elastic}. As shown in Fig.~\ref{fig: concrete_defects}, these defects often occur when concrete stops flowing prematurely due to factors such as thixotropy, where the material exhibits a time-dependent increase in structural strength, as well as interactions with obstacles such as reinforcement bars, molds, support fluids, and surrounding materials \citep{sun2024pile, mahjoubi2025thermal}. Investigating the rheological properties of fresh concrete, including its workability and the mechanics behind flow interruption, is essential for developing strategies to minimize defects, optimize casting processes, and enhance the overall performance of concrete infrastructures \citep{wilkes2024practical}. Furthermore, optimizing concrete mix design to improve flow properties can significantly reduce cement consumption and enhance material efficiency, contributing to lower carbon emissions and promoting greener construction practices \citep{nilimaa2023smart, barbhuiya2024decarbonising}. 

Various mechanistic models have been proposed to describe the rheological properties of fresh concrete, with the Bingham model \citep{bingham1922fluidity} being the most widely adopted due to its simplicity and practical applicability \citep{deeb20143d, li2021numerical, yu2024modeling}. The Bingham model treats fresh concrete as a yield-stress fluid (often referred to as a rigid-plastic fluid), assuming that the material behaves as a rigid body when the applied shear stress is below the yield stress threshold, $\tau_0$. Once the applied stress exceeds this threshold, the material begins to flow, and the relationship between shear stress and shear rate is taken to be linear. This behavior can be expressed in terms of the shear rate as:
\begin{equation}
\dot{\gamma} =
\begin{cases}
0, & \tau < \tau_0\,,\\
\dfrac{\tau - \tau_0}{\eta}, & \tau \ge \tau_0\,.
\end{cases}
\label{eq:bingham}
\end{equation}
where $\tau_0$ represents the critical stress required to initiate flow, while the plastic viscosity $\eta$ governs the post-yield flow behavior.

Another commonly used model is the power-law fluid \citep{rao2004peristaltic, manica2004simulation, pichler2017apparent}, which can describe shear-thinning or shear-thickening behavior but lacks the ability to account for yield stress, making it inadequate for capturing the initiation and cessation of flow in fresh concrete. Meanwhile, the Herschel–Bulkley (HB) model \citep{herschel1926konsistenzmessungen} extends the Bingham model’s capabilities and provides a more comprehensive representation of rheological behavior by incorporating a nonlinear shear-stress--strain-rate relationship in different ranges of shear conditions \citep{cai2025cfd, li2021numerical}.

The aforementioned rheological models, in their standard form, often fail to adequately account for time-dependent effects in fresh concrete. Among those, thixotropy stands out as a key property, characterized by a reduction in viscosity under shear and a gradual recovery when the material is at rest. Thixotropy significantly influences concrete flowability during practical operations such as formwork filling, staged casting with sufficient time gaps between pours, or flow through congested reinforcement. In these scenarios, flocculation can restrict flow, potentially causing stoppages and leading to defects that compromise the quality of the final structure. \cite{roussel2006} introduced a foundational framework that employs simplified mathematical formulations to quantify the kinetics of thixotropic behavior. In their model, a constant thixotropy coefficient $A_{\mathrm{thix}}$ is introduced to characterize the flocculation (i.e.~structural build-up) rate of a concrete mixture, and the deflocculation (i.e.~structural breakdown) term is expressed as a linear relationship of shear rate. This formulation, however, treats the effects of physical particle interactions and chemical hydration on the overall structural build-up as a reversible aggregated contribution, which remains a simplification and modeling assumption \citep{jiao2021thixotropic}.

Building on Roussel's model, subsequent refined thixotropic models have been proposed to address its limitations and provide a more comprehensive description of the structural evolution of cementitious materials. These models generally: (i) incorporate the irreversible contribution of chemical hydration, thereby decoupling it from reversible flocculation processes \citep{perrot2016structural, lecompte2017non}; (ii) consider the influence of multiple factors, such as the surface coverage of superplasticizer molecules, the packing density of cementitious particles, and the volume fraction of aggregate, on the flocculation rate \citep{mahaut2008effect, lowke2018thixotropy}; and \hl{(iii) employ nonlinear laws to describe the flocculation and deflocculation process, enabling the prediction of more complex structural build-up and breakdown behaviors \citep{wallevik2009rheological, tao2024modeling, teng2020effect, ma2018experimental}. However, these advanced models typically require a greater number of input parameters and state variables, increasing their complexity for calibration and thereby limiting their practical usability in engineering industries}. Despite these advancements, \cite{roussel2006}'s thixotropy model remains the most widely used in practice due to its ease of implementation and reduced modeling complexity \citep{de2018cfd, wilkes2023investigating}.

Integrating rheological models with numerical simulation methods provides powerful tools for understanding and predicting the time-dependent flow behavior of fresh concrete under practical casting conditions. Among various numerical approaches, the mesh-based Finite Volume Method (FVM) is the most widely adopted within the concrete community. Since FVM is inherently Eulerian, it typically relies on ad hoc interface tracking techniques, such as the Volume of Fluid (VOF) method, to capture free-surface boundaries \citep{cai2025cfd, comminal2018numerical, comminal2020modelling, shin2023flow}. Through OpenFOAM \citep{jasak2007openfoam} with VOF, concrete has been modeled as a Bingham fluid \citep{mitchell25} with the intent of highlighting the origin of defects related to bulk flow behaviors. However, FVM faces fundamental challenges in simulating history-dependent material behavior, such as fresh concrete with evolving thixotropic properties, due to the difficulty of tracking material state variables as they advect across a fixed grid. This, therefore, requires additional mapping procedures to properly update and transport these variables over time.

In contrast, Lagrangian-based methods, such as Smoothed Particle Hydrodynamics (SPH) \citep{deeb20143d, cao2017numerical, ouyang2022hybrid, yu2024modeling}, the Lattice Boltzmann Method (LBM) \citep{li2021numerical}, and the Discrete Element Method (DEM) \citep{wu2023discrete, mu2023research}, track material motion through continuum or discrete particle representations, making them naturally suited for free-surface flows and large-deformation simulations. While these methods eliminate the need for explicit interface tracking, techniques like SPH and DEM require neighbor-search operations, which are known to be computationally demanding. Furthermore, DEM is particularly effective for representing interactions between concrete aggregates. However, because it relies solely on particle–particle contact, DEM encounters difficulties in capturing the continuous flow of cement paste, which follows a non-Newtonian constitutive response as finer grains mix with water. This limitation often necessitates coupling DEM with a continuum solver to represent cement paste–aggregate interactions, thereby increasing model complexity and computational cost \citep{nan2021clogging, ding2025study}. The computational cost of DEM becomes even more significant when non-spherical aggregate geometries are considered, although recent advances in GPU computing offer promising opportunities to mitigate these challenges.

Hybrid mesh-particle methods, like the Particle Finite Element Method (PFEM) \citep{reinold2022} and the Material Point Method (MPM) \citep{wilkes2023investigating, yildizdag2024numerical}, leverage the advantages of both Lagrangian particles and finite element grids, rendering them effective and accurate in simulating large deformations and free surface flows. Despite all its benefits, PFEM is generally computationally demanding as it requires frequent remeshing. By contrast, MPM employs a fixed Eulerian background grid coupled with Lagrangian material points, thereby avoiding expensive neighbor searches and remeshing while concurrently capturing history-dependent material behavior and extreme deformations. Owing to these numerical benefits, MPM has been widely applied to simulate many large-deformation problems in solids and fluids \citep{zhang2016material, soga2016, de2020material, yu2024multiscale, yu2024thermo}, including other types of viscoplastic materials, such as foams \citep{yue2015continuum}.

Modeling the flow behavior of fresh concrete has predominantly relied on treating it as a non-Newtonian single-phase fluid. However, this assumption presents significant limitations. A key drawback is that such models inherently depend on a non-zero shear strain rate to sustain shear stress, implying that the material must be continuously flowing to support shear. As a result, the material would not stop flowing unless the shear strain rate approaches zero \citep{xu2021numerical}, which is only achieved under a hydrostatic state, where the free surface becomes perpendicular to the direction of gravity. In casting simulations, this leads to unrealistic predictions in which, as time approaches infinity, the entire casting domain would eventually fill without defects. These limitations become even more critical with the increasing adoption of automated construction technologies, particularly 3D concrete printing, where the deposited material must rapidly transition from a flowing to a load-bearing state while maintaining its geometric stability. Accurate prediction of buildability, shape retention, and interlayer stability therefore requires a constitutive framework that can capture both fluid-like flow and solid-like mechanical response within a unified yet clearly distinguishable formulation.

\hl{Some previous studies have sought to address this limitation by introducing empirical flow-stopping criteria, for example by specifying a critical velocity \citep{yu2024modeling} or a critical shear rate \citep{mohammad20222d} below which the material is assumed to come to rest. Alternative approaches compare the yield stress with the tangential resistance at the boundary \citep{xu2021applicability} to identify when flow ceases. Although these methods enhance the numerical prediction of flow arrest, they depend on problem-specific threshold parameters that lack a rigorous constitutive foundation and must be determined through calibration. In addition, non-Newtonian fluid models often assume numerical stabilization techniques, such as Papanastasiou regularization \citep{papanastasiou1987flows}, to address stress discontinuities as the shear rate approaches zero. While this regularization effectively enhances numerical stability, it also introduces an artificial parameter. Raising the value of this parameter yields a better approximation of the ideal Bingham behavior, but at the cost of producing unrealistically high apparent viscosities at low shear rates. This, in turn, increases numerical stiffness and, in certain cases, causes kinematic locking, reducing computational stability and overall efficiency.}

To address these limitations, this study adopts a solid-mechanics modeling approach by formulating the rheology of fresh concrete within an elasto-viscoplastic framework. Inspired by the work of \citet{dunatunga2015continuum} for modeling dry granular materials, this work incorporates a rate-dependent yield condition, allowing the material to behave elastically below the yield surface and to exhibit a rate-dependent inelastic response once yielding occurs. As a result, the simulated flow naturally stops when stresses fall below the yield threshold, enabling a realistic transition to a static state \citep{reinold2022}. Moreover, the model eliminates the need for artificial stabilization techniques, providing a more physically consistent parameterization of fresh concrete behavior. Building on Roussel’s pioneering work \citep{roussel2006}, a thixotropy evolution equation is also integrated into the proposed framework. Incorporating thixotropic mechanisms accounts for flocculation and deflocculation processes, enabling a more comprehensive representation of structural build-up over time and breakdown during flow. The proposed framework is implemented within the MPM, providing a robust and versatile computational approach for simulating fresh concrete flow under various conditions with history-dependent variables. Overall, this study aims to overcome the inherent limitations of fluid-based models and to offer a more physically consistent and predictive tool for modeling concrete processes with flow-rest transitions.

The remaining part of this paper is structured as follows. Section \ref{sec: sec2} introduces the continuum theory underlying the proposed elasto-viscoplastic model. Section \ref{sec: sec3} elaborates on the modeling framework, including a brief discussion of its implementation within the MPM and the stress update algorithm. Section \ref{sec: Numerical examples} provides numerical examples to validate the proposed model and demonstrate its application and limitations in simulating fresh concrete flow under various conditions. Finally, Section \ref{sec: conclusion} gives some concluding remarks and discusses potential directions for future research.

\section{Theoretical framework}
\label{sec: sec2}

In this section, we present the theoretical framework for modeling fresh concrete, including the governing equations and associated constitutive assumptions. We begin by introducing the mathematical notation used throughout this work. The symbols $\dot{\square}$ and $\ddot{\square}$ denote first- and second-order material time derivatives, respectively. The operators $\square\cdot\square$ and $\square:\square$ represent the single and double tensor contractions, and $\square\otimes\square$ denotes the tensor product. For an arbitrary second-order tensor $\tb A$, its trace is defined as $\mathrm{tr}(\tb A)$, and its deviatoric part is given by $\tb A^* := \tb A - \mathrm{tr}(\tb A)/3$. The symmetrization and skew-symmetrization operators are expressed as $\mathrm{sym}(\tb A) := (\tb A + \tb A^T)/2$ and $\mathrm{skw}(\tb A) := (\tb A - \tb A^T)/2$, respectively. Throughout this study, bold or blackboard bold notation is used to denote spatial variables in tensorial form.

\subsection{Governing equations}
The conservation laws of mass and momentum govern the flow of fresh concrete. In the Lagrangian description, they are given as follows,
\begin{gather}
    \dot\rho+\rho\nabla\cdot\boldsymbol{v}=0\,, \\
    \rho\dot{\boldsymbol{v}}-\nabla\cdot\boldsymbol{\sigma}-\rho\boldsymbol{b}=\boldsymbol{0}\,,
    \label{eq:momentum_bal}
\end{gather}
where $\rho$ is the bulk mass density, $\boldsymbol{v}$ is the velocity field, $\boldsymbol{\sigma}$ is the Cauchy stress tensor, and $\boldsymbol{b}$ is the body force vector. In a boundary value problem (BVP), the objective is to solve the balance equations above with time range $t\in[0,T]$ subjected to the following Dirichlet and Neumann boundary conditions:
\begin{subequations}
\begin{align}
\boldsymbol{v}&=\hat{\boldsymbol{v}}\qquad \mathrm{on}\quad \Gamma_v,\\
\boldsymbol{\sigma}\cdot\boldsymbol{n}&=\hat{\boldsymbol{t}}\qquad\mathrm{on}\quad \Gamma_{t},
\end{align}   
\end{subequations}
where $\boldsymbol{n}$ is the outward unit normal vector, whereas $\hat{\boldsymbol{v}}$ and $\hat{\boldsymbol{t}}$ denote the prescribed boundary velocity and traction acting on the corresponding surfaces $\Gamma_v$ and $\Gamma_t$, respectively.


\subsection{Constitutive theory}

Fresh concrete is considered to be homogenized into a single-phase continuum body that follows Bingham rheology (cf.~Eq.~\eqref{eq:bingham}), where the material flows with a constant viscosity $\eta$ once it is sheared beyond a certain yield stress $\tau_0$ ($\eta$ is often referred to as the plastic viscosity). Since concrete is a thixotropic material, its microscopic constituents flocculate at rest and deflocculate when shearing; a thixotropy-dependent variation of the yield stress should be incorporated to accurately capture its flow behavior. Instead of treating fresh concrete as a rigid–plastic fluid, in which the material is assumed rigid when the shear stress is below $\tau_0$, the present work models it as an elasto-viscoplastic solid. In this framework, when the stress state lies below the yield surface $\tau_0$, the material behaves elastically and can sustain stress. This simple shift in paradigm allows the physical cessation of material flow to be simulated, which cannot be represented by fluid models and often leads to unphysical long-term flow predictions. Further discussions comparing these two modeling approaches are provided in the subsequent section.

The total stress of the material can be decomposed into the isotropic (or volumetric) pressure and the deviatoric stress as
\begin{equation}
    \boldsymbol{\sigma}=\boldsymbol{\tau}-p\boldsymbol I,   
\end{equation}
where $p=-\mathrm{tr}(\tb \sigma)/3$ is the mean pressure (positive in compression), $\boldsymbol{\tau}$ is the shear stress tensor (positive in tension), and $\boldsymbol{I}$ is the second-order identity tensor.

The fresh concrete is assumed to behave as a weakly compressible fluid \citep{yu2024modeling}. Its volumetric response can be described by the following equation of state:
\begin{equation}
p = \frac{\rho_0 c_0^2}{\gamma}\left[\exp\left(-\gamma \varepsilon_v \right)-1\right]\,,
\label{eq: EOS1}
\end{equation}
which is a slight rewriting of the Tait–Murnaghan equation of state (see \citet{macdonald1966some}). Here, $\rho_0$ is the initial mass density, and $\gamma$ is a material constant, typically chosen as $\gamma=7$ to reproduce weakly compressible fluid behavior. The speed of sound in the medium is denoted as $c_0 = \sqrt{K_0 / \rho_0}$, where $K_0$ is the initial bulk modulus of fresh concrete.

The volumetric strain $\varepsilon_v$ is obtained by integrating the volumetric strain rate:
\begin{eqnarray}
    \varepsilon_v = \int \mathrm{tr}(\tb D) \td t\,,
\end{eqnarray}
where $\tb D$ is the strain-rate tensor, defined as the symmetric part of the velocity-gradient tensor $\tb L$, i.e.~$\tb D = \mathrm{sym}(\tb L)$ and $\tb L = \nabla \tb v$.

When flowing, the shear stress response of fresh concrete follows a generalized Newtonian fluid model \citep{bird1987dynamics, chhabra2025non}, expressed as
\begin{equation}
    \boldsymbol{\tau} = 2\eta_{a}\boldsymbol{D}^*\,, \qquad \mathrm{with} \qquad \eta_{a}=\frac{\tau}{\dot\gamma}\,.
\end{equation}
Here, $\boldsymbol{D}^*$ is the deviatoric strain-rate tensor, whose norm is defined as $\dot{\gamma} = \sqrt{2\tb D^{*} : \tb D^{*}}$. The quantity $\eta_{a}$ denotes the apparent viscosity, and $\tau = \sqrt{J_2(\tb\sigma)} = \sqrt{(\tb\sigma^* : \tb\sigma^*)/2}$ is the second invariant of the deviatoric stress tensor.

\subsection{Plastic response and rheology}

In the following subsections, we will elaborate further on how the deviatoric response of Bingham rheology is modeled under two frameworks: (i) the rigid–plastic fluid (with suitable regularization) and (ii) the elasto-viscoplastic solid. We also discuss how concrete thixotropy is incorporated into these models.

\subsubsection{Rigid-plastic Bingham fluid with regularization}
\label{sec:PR-bingham model}

Fresh concrete modeled as a Bingham fluid behaves as a rigid body before yielding and as a viscous material after yielding. However, modeling concrete as a yielding fluid is constitutively challenging, as it still requires approximating the onset of flow when $\tau > \tau_0$. To represent the behavior before and after yield, a commonly adopted modification of the Bingham model is the bi-viscous model \citep{o1984numerical}, which introduces a critical shear rate $\dot{\gamma}_c$ to define the transition between the rigid and viscous regimes. The bi-viscous model is written as
\begin{equation}
    \boldsymbol{\tau} = 
    \begin{cases} 
        2\eta_0\boldsymbol{D}^*\,,  & \text{if } |\dot{\gamma}| \leq \dot{\gamma}_c\,, \\ 
        2\left(\eta + \frac{\tau_0}{\dot{\gamma}}\right)\boldsymbol{D}^*\,, & \text{otherwise}\,,
    \end{cases}
\label{eq:tau_ij}
\end{equation}
where $\eta_0$ is the pre-yield viscosity. In practice, $\eta_0$ is assigned a very large value to approximate the near-rigid behavior of the material in the unyielded region, effectively limiting strain rates before yielding occurs. The bi-viscous model provides a distinction between pre-yield and post-yield regions through the critical shear rate $\dot{\gamma}_c$; however, the abrupt discontinuity at this transition, together with the excessively large $\eta_0$, often presents a computational challenge.

\begin{figure}[h!]
    \centering
    \includegraphics[width=0.55\linewidth]{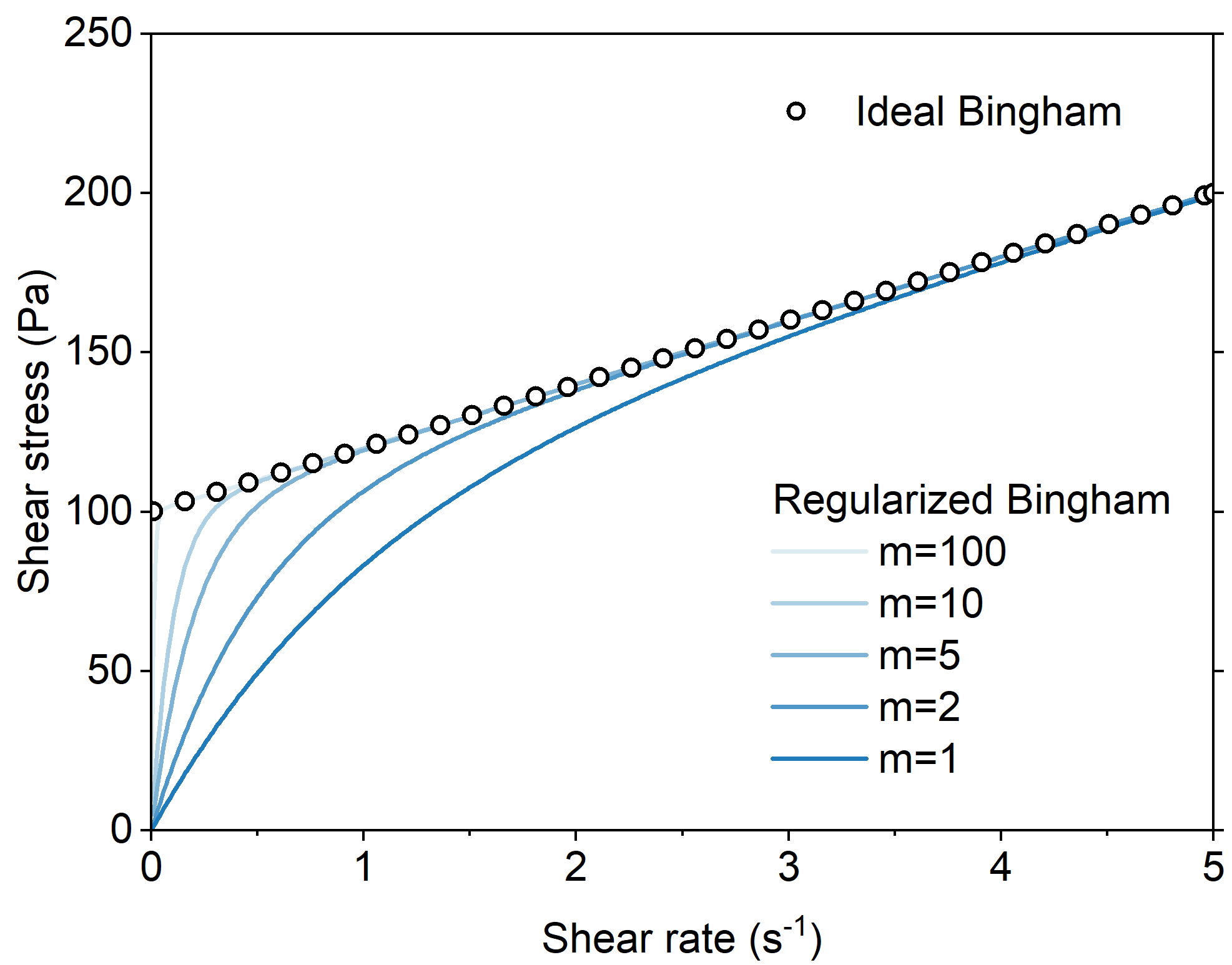}
    \caption{Stress-strain-rate relationship of the regularized Bingham model.}
    \label{regularized_binghaml}
\end{figure}

An alternative regularization method, which is probably the most commonly used in recent studies, involves adopting a continuous model that eliminates the distinct discontinuity between the rigid and plastic regions. The Papanastasiou \citep{papanastasiou1987flows} model is the most widely used among these continuous regularizations. The exponential regularization of the Bingham model, often referred to as the Bingham–Papanastasiou model, introduces a parameter $m$ such that the shear stress is given by:
\begin{equation}
    \boldsymbol{\tau} = 2 \left[ \eta + \frac{\tau_0}{\dot{\gamma}} (1 - e^{-m \dot{\gamma}}) \right] \boldsymbol{D}^*\,.
  \label{eq:regularized_bingham}
\end{equation}
Here, the parameter $m$ denotes the stress growth (or regularization) parameter, which controls how rapidly the material response transitions from the pre-yield to the post-yield regime. A larger $m$ value produces a steeper transition and more closely approximates the ideal Bingham behavior. The effect of varying $m$ is illustrated in Fig.~\ref{regularized_binghaml}, where increasing $m$ leads to higher apparent viscosity at low shear rates.

While the Bingham–Papanastasiou model resolves the issue of stress discontinuity, it introduces another theoretical limitation. In this formulation, the material can only sustain shear stress when it is flowing, i.e.~$\tau>0$ if and only if $\dot{\gamma}>0$. As a result, the material continues to flow indefinitely under any nonzero shear stress and can only come to rest when the shear stress vanishes, such as under lithostatic equilibrium in a gravity-driven flow. In other words, fresh concrete materials described by this model would theoretically keep flowing over time, gradually flattening until their free surface becomes perpendicular to the direction of gravity. This ``perpetual flow" behavior is physically unrealistic, since real fresh concrete can cease flowing and retain its shape once internal restructuring and thixotropic effects develop upon rest (cf.~Fig.~\ref{fig: concrete_defects}). This fundamental limitation motivates the adoption of an elasto-viscoplastic solid framework, enabling a physically realistic transition to a resting state that is particularly important for long-period concrete casting simulations.

\subsubsection{Thixotropic effect}
To capture this transition to rest in a physically meaningful manner, we next introduce a thixotropy model that accounts for the time-dependent evolution of the internal structure of fresh concrete. In this work, we incorporate the thixotropy model for fresh concrete proposed by \citet{roussel2006}. This model introduces a flocculation state parameter, $\lambda$, which evolves over time such that the effective yield stress is expressed as
\begin{equation}
    \Tilde{\tau}_{0}(t) = (1 + \lambda(t)) \tau_{0}\,,
    \label{eq:dynamic_tau}
\end{equation}
where $\tau_0$ is the reference yield stress, measured without resting time or right after mixing. 

\begin{figure}[h!]
    \centering
    \includegraphics[width=0.8\linewidth]{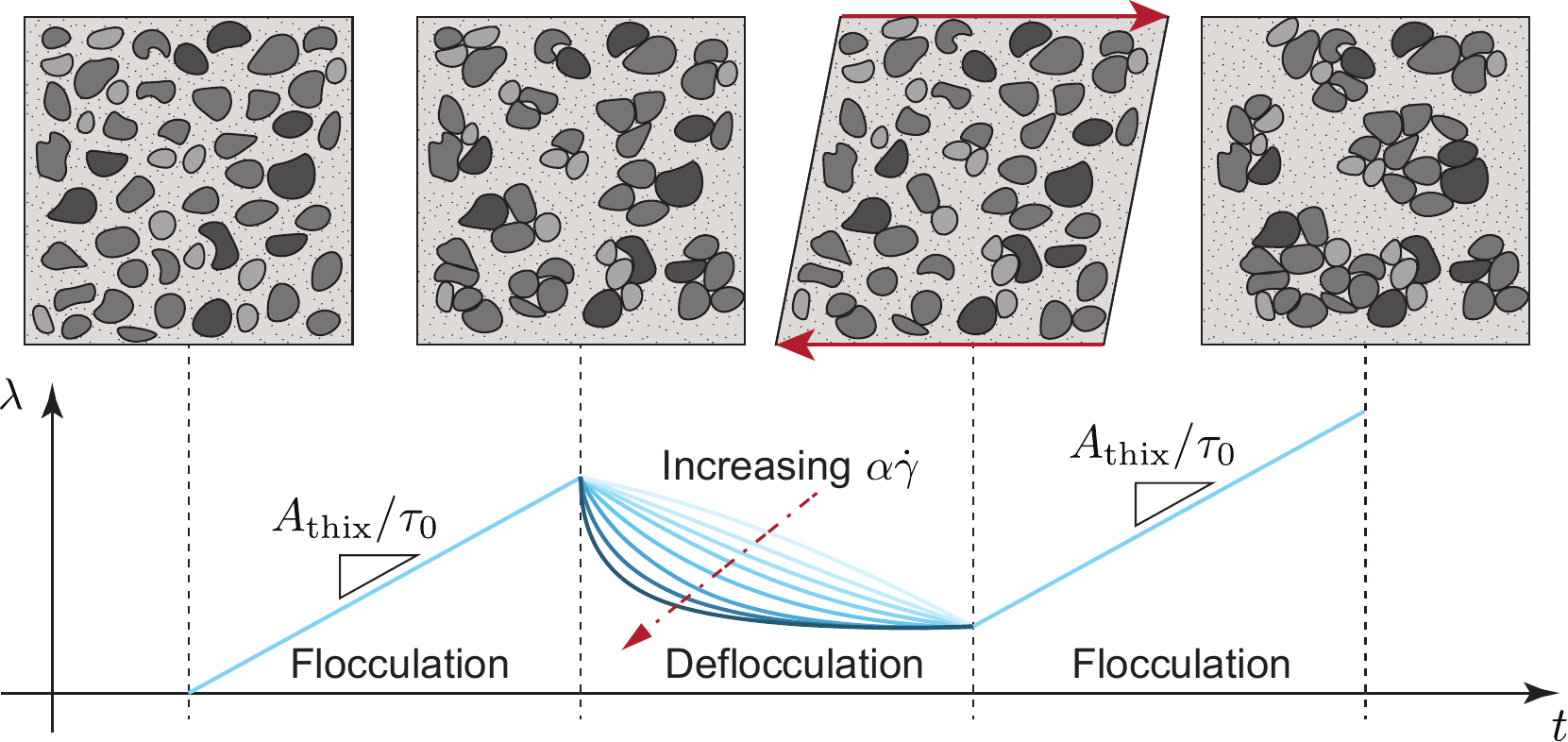}
    \caption{Schematic illustration of flocculation- and deflocculation-induced strengthening in fresh concrete characterized through the evolution of $\lambda$ over time.}
    \label{fig:flocculation}
\end{figure}

The parameter $\lambda$ characterizes the internal structural state of the material, depending on both the rest period and the experienced shear rate. When the material is at rest, $\lambda$ increases due to particle flocculation, whereas when the material is sheared, $\lambda$ decreases as a result of deflocculation (see Fig.~\ref{fig:flocculation}). Flocculation increases the effective yield stress $\tilde{\tau}_0$ because the formation of flocculated clusters enhances shearing resistance. The rate of change of $\lambda$ can be described, following Roussel’s observations, as
\begin{equation}
    \frac{\partial \lambda}{\partial t} = \frac{A_{\mathrm{thix}}}{\tau_0} - \alpha \lambda \dot{\gamma}\,,
    \label{eq:lambda_rate}
\end{equation}
where $A_{\mathrm{thix}}$ is the flocculation rate (in Pa/s) and $\alpha$ is the deflocculation coefficient. Concrete with $A_{\mathrm{thix}} < 0.1$ Pa/s is typically classified as non-thixotropic, whereas concrete with $A_{\mathrm{thix}} > 0.5$ Pa/s is considered highly thixotropic. While the present formulation assumes a constant $A_{\mathrm{thix}}$, previous studies have reported that its value may depend on the apparent shear rate \citep{wallevik2015avoiding, thiedeitz2020box}. 

Furthermore, the value of $\alpha$ also plays a critical role, as it governs the exponential decrease of shear stress under a constant shear rate, with a characteristic deflocculation time scale given by $1/(\alpha \dot{\gamma})$. Its magnitude depends on the number and strength of interparticle links within the fresh concrete \citep{assaad2003assessment}. Reported values in the literature span several orders of magnitude, typically ranging from $10^{-3}$ to $10^{-1}$ \citep{assaad2003assessment, roussel2005steady, roussel2006, qian2016flow, Thiedeitz2024ThixotropicPastes}.

By combining Eqs.~\eqref{eq:regularized_bingham}, \eqref{eq:dynamic_tau}, and \eqref{eq:lambda_rate}, we obtain the thixotropic Bingham fluid model, referred to as the Papanastasiou–Roussel Bingham (PR-Bingham) model \citep{wilkes2023investigating}, expressed as
\begin{equation}
\boldsymbol{\tau} = 2 \left[ \eta + \frac{\tau_0}{\dot{\gamma}}(1 + \lambda) (1 - e^{-m \dot{\gamma}}) \right] \boldsymbol{D}^*\,.
\end{equation}
    
\subsubsection{Elasto-viscoplasticity Bingham solid with thixotropy}
\label{sec:solid_model}

To address the limitations of the Bingham fluid model --- specifically, its inability to sustain shear stress in a non-flowing state --- we model fresh concrete as an elasto-viscoplastic solid following the continuum mechanics framework \citep{steigmann2023course}. In this model, the material behaves as a nonlinear elastic solid in shear prior to yielding, with the stress state remaining below the yield threshold. Once yielding occurs, the material transitions to a viscoplastic solid, exhibiting rate-dependent dissipation characterized by the plastic viscosity $\eta$. This formulation inherently provides a finite stopping criterion: when the driving stress decreases below the yield threshold, the response becomes purely elastic, and inelastic deformation ceases, thereby naturally arresting the material flow.

Central to this formulation, and distinguishing it from the fluid formulation, is the additive decomposition of the strain-rate tensor $\tb D$, which is expressed as the sum of elastic and plastic components, $\tb D^e$ and $\tb D^p$, respectively, i.e.~$\tb D = \tb D^e + \tb D^p$. The elastic strain-rate $\tb D^e$ represents the recoverable portion of the total strain rate that builds up elastic strain energy in the material; this strain is released once the loading is removed, resulting in elastic recovery. In contrast, the plastic strain-rate $\tb D^p$ corresponds to the irrecoverable component of deformation, which accumulates over time and produces permanent, or inelastic, deformation of the material. Similarly, the spin-rate tensor $\tb W = \skw{\tb L}$ can also be decomposed additively into its elastic and plastic parts, i.e.~$\tb W = \tb W^e + \tb W^p$.

We define a yield surface $f$ \hl{analogous} to the von Mises yield criterion \hl{but extended to incorporate the Bingham rheology, as}
\hl{
\begin{eqnarray}
    f = \tau - \tau_y\,, 
    \label{eq:von_misses}
\end{eqnarray}
where $\tau_y$ is the thixotropic- and rate-dependent yield stress, defined as:
\begin{eqnarray}
\tau_y:=\widehat{\tau}_y(\lambda,\dot{\gamma}^p)=(1+\lambda){\tau}_0 +\eta \dot{\gamma}^p\,.
\label{eq:rate_dependent_yield_stress}
\end{eqnarray}
During plastic flow,} the yield function remains constant ($f = 0$), ensuring that the stress state always lies on the yield surface; this is known as Prager’s consistency condition. The plastically admissible stresses are those within the elastic domain, either inside or on the boundary of the yield surface, that satisfy the following Karush–Kuhn–Tucker (KKT) conditions:
\begin{gather}
    \dot{\gamma}^p f=0\,, \quad \dot{\gamma}^p\geq 0\,, \quad f\leq 0\,.\label{eq:yield_cond}
\end{gather}
\hl{In the above expressions,} $\dot{\gamma}^p$ denotes the equivalent plastic shear rate; this will be discussed shortly.

When the stress state lies below the yield surface ($f < 0$ \hl{and $\dot{\gamma}^p=0$}), the material is considered to be in a solid-like state, and the stress response is governed by a nonlinear elastic model. \hl{Here, we assume that elastic deformation remains small (under the rigid-plastic limit assumption), yet not strictly zero, in comparison to plastic deformation, i.e.~$\tb D^p\gg\tb D^e$, so that $\tb D\approx\tb D^p$. This assumption is supported by the fact that elastic stiffness is relatively high compared with the stress states at which yielding occurs, implying that post-yield deformation is predominantly governed by plastic flow \citep{steigmann2023course}.} To ensure stress objectivity under superposed rigid motion, an objective stress rate must be employed; in this study, we adopt the Jaumann rate, i.e.,
\begin{eqnarray}
    \overset{\nabla}{\tb \sigma} \equiv \dot{\tb \sigma} - \tb{W} \tb \sigma + \tb \sigma \tb{W}=\mathbb{C}:\tb D^e\,,
\end{eqnarray}
where the fourth-order elastic constitutive tensor is defined as
\begin{eqnarray}
    \mathbb C= K \tb I \otimes \tb I + 2 G \left( \mathbb{I}-\frac{1}{3} \tb I \otimes \tb I \right)\,,
    \label{eq:elastic_tangent_matrix}
\end{eqnarray}
with $K$ and $G$ denoting the bulk and shear moduli, respectively, and $\mathbb{I}$ representing the fourth-order identity tensor. The bulk modulus $K$ and $G$ can be related to the weakly compressible fluid model defined in Eq.~\eqref{eq: EOS1} through:
\begin{eqnarray}
    K = -\frac{\pd p}{\pd \varepsilon_v} = \rho_0 c_0^2\exp\left(-\gamma \varepsilon_v \right)\,, \qquad G=\frac{3K (1-2\nu)}{2(1+\nu)}\,.
\end{eqnarray}
For fresh concrete, Poisson's ratio $\nu$ typically ranges between 0.45 and 0.49, indicating its nearly incompressible nature.

Upon yielding \hl{(when $f=0$)}, the plastic flow rule defines the plastic strain-rate tensor $\tb D^p$ as proportional to the normalized deviatoric stress direction, scaled by the equivalent plastic shear rate $\dot{\gamma}^p$:
\begin{equation}
    \tb D^p = \dot{\gamma}^p \frac{\tb\tau}{2\tau}\,.
    \label{eq:flow_rule}
\end{equation}
Note that the flow rule is isochoric, $\mathrm{tr}(\tb D^p) = 0$; hence, $\tb D^p = \tb D^{p*}$ and $\dot{\gamma}^p = \sqrt{2\,\tb D^{p} : \tb D^{p}}$. In the plastic state, where $\dot{\gamma}^p$ is nonzero, \hl{in accordance with the KKT condition (Eq.~\eqref{eq:yield_cond}), the shear stress $\tau$ follows the thixotropic Bingham rheology and is equal to the rate-dependent yield stress $\tau_y$, expressed as:
\begin{eqnarray}
    \tau = \tau_y= (1+\lambda)\tau_0+\eta \dot{\gamma}^p\,.
    \label{eq:plastic_shear_stress}
\end{eqnarray}
Upon substituting Eq.~\eqref{eq:plastic_shear_stress} into \eqref{eq:flow_rule}} and rearranging the terms, the deviatoric stress tensor can be expressed in terms of the plastic strain rates as:
\begin{equation}
\boldsymbol{\tau} = 2 \left[ \eta + \frac{\tau_0}{\dot{\gamma}^p} \left(1 + \lambda\right) \right] \boldsymbol{D}^p\,.
\end{equation}
\hl{Furthermore, following rigid plastic limit assumption}, the evolution of the flocculation state parameter $\lambda$ (Eq.~\eqref{eq:lambda_rate}) can be rewritten in terms of $\dot{\gamma}^p$ as:
\begin{equation}
    \frac{\pd \lambda}{\pd t} = \frac{A_{\mathrm{thix}}}{\tau_0} - \alpha \lambda \dot{\gamma}^p\,.
    \label{eq:lambda_rate_plastic}
\end{equation}

By integrating elasticity, rate-dependent plasticity, and thixotropy, the elasto-viscoplastic solid model eliminates the need for ad hoc stopping criteria \citep{yu2024modeling, mohammad20222d, xu2021applicability} and additional regularization methods. Instead, flow cessation emerges naturally when the stress state falls below the yield threshold, allowing elastic recovery to dominate while the inelastic strain rate vanishes ($||\boldsymbol{D}^p|| \rightarrow 0$). Meanwhile, the thixotropic term gradually increases the yield stress over time, making the material progressively more resistant to yielding and thereby aligning the framework with experimentally observed fluid-solid transitions in fresh concrete flow.



\section{Modeling framework and discretization}
\label{sec: sec3}

\subsection{Material point discretization}

\begin{figure}[h!]
    \centering
    \includegraphics[width=\linewidth]{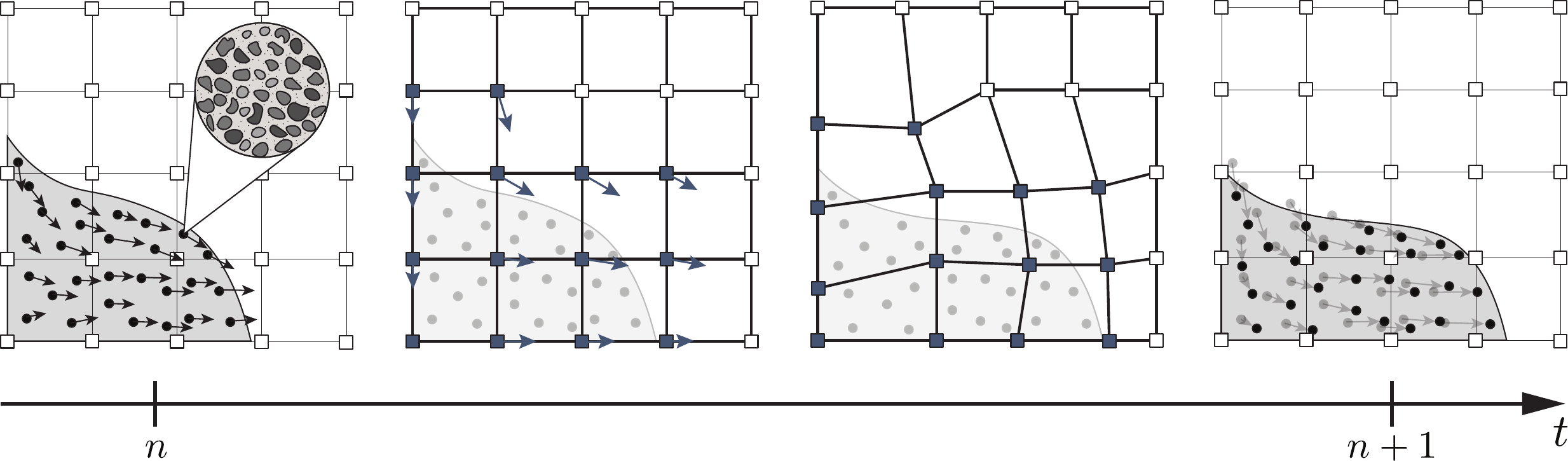}
    \caption{MPM computational cycle at each time step. In the MPM, a continuum body is discretized into a set of material points that move spatially. A material point is not a real physical particle, but rather a control volume representing a homogenized portion of the material domain. At the beginning of each time step, the kinematic fields are mapped from the material points to a background grid (with active nodes highlighted in blue). The next step is to solve the mass and momentum balance equations, considering imposed boundary conditions. At the end of the step, the kinematic fields are transferred back to the material points, where they are then advected through space. After deformation, the background grid can be reset to its original shape.}
    \label{fig:mpm}
\end{figure}

In this work, we employ the Material Point Method (MPM) \citep{Sulsky1994}, a continuum simulation technique derived from the Fluid–Implicit–Particle (FLIP) method \citep{brackbill1986flip} for modeling large-deformation continua. In MPM, the body is discretized into Lagrangian material points that store history-dependent variables of the continuum, while a background Eulerian grid is used to solve the governing equations of motion. Because all state information is carried by the material points, the background mesh can be reset at the beginning of each time step without any loss of information, allowing the method to handle large deformations without introducing errors associated with mesh distortion. A schematic of the MPM cycle is shown in Fig.~\ref{fig:mpm}. 

The MPM framework follows a variational formulation based on the Bubnov–Galerkin method. The weak form of the momentum balance equation (Eq.~\eqref{eq:momentum_bal}) can be written as:
\begin{equation}
    \int_{\Omega} \delta\boldsymbol{u}\cdot \rho \ddot{\boldsymbol{u}} \, d\Omega + \int_{\Omega} \nabla \delta\boldsymbol{u}: \boldsymbol{\sigma}  \, d\Omega - \int_{\Omega} \delta\boldsymbol{u}\cdot \rho\boldsymbol{b} \, d\Omega - \int_{\Gamma_t} \delta\boldsymbol{u}\cdot \bar{\boldsymbol{t}} \, d\Gamma = {0}\,,
    \label{eq: mom_weak_form}   
\end{equation}
where $\delta \tb u$ denotes an arbitrary test function. Using finite element discretization, the displacement field and its corresponding test function are approximated as
\begin{equation}
    \tb u (\tb x) = \sum_{I=1}^{n_n} S_I(\tb x) \tb u_I\,, \qquad \delta \tb u (\tb x) = \sum_{I=1}^{n_n} S_I(\tb x) \delta \tb u_I\,,
\end{equation}
where $n_n$ is the total number of nodes and $S_I$ is the shape function associated with node $I$ in the background grid. Employing this approach and invoking the arbitrariness of the test function, we can discretize the momentum balance equation and solve for the nodal acceleration $\tb{a}_J^{n+1}$ at time $t=n+1$ as:
\begin{equation}
    m_{I}^n \tb{a}_I^{n+1} = \tb{f}_I^{\mathrm{int},n} + \tb{f}_I^{\mathrm{ext},n}\,,
    \label{eq:discrete_mom}
\end{equation}
where the nodal mass $m_I$, internal force $\tb{f}_I^{\mathrm{int}}$, and external force $\tb{f}_{I}^{\mathrm{ext}}$ are defined as:
\begin{equation}
    m_{I}^n = \sum_{p=1}^{n_p} S_I (\tb x_p) m_p\,,
    \label{eq:mpm-mass-matrix}
\end{equation}
\begin{equation}
    \tb{f}_I^{\mathrm{int},n} = - \sum_{p=1}^{n_p} \nabla S_I (\tb x_p) \cdot {\pmb\sigma}^n_p V_p^{n+1}\,,
    \label{eq:mpm-internal-force}
\end{equation}
\begin{equation}
    \tb{f}_I^{\mathrm{ext},n} = \sum_{p=1}^{n_p} S_I (\tb x_p) \tb{b}_p m_p + \int_{ \Gamma_t} S_{I}(\tb x) \bar{\boldsymbol{t}}(\tb{x}) \td \Gamma\,.
    \label{eq:mpm-external-force}
\end{equation}
In the notation above, the variables at grid nodes are indicated with subscript $I$, whereas subscript $p$ denotes material point variables. The total number of material points located within the local neighborhood of $I$ is indicated by $n_p$. Here, $m_p$ and $V_p$ denote the material point mass and volume, respectively. Notice that the traction surface integral in Eq.~\eqref{eq:mpm-external-force} is left in its weak form for now, as its discretization is non-trivial due to boundary nonconformity; see \citep{bing2019b,chandra2021nonconforming, liang2023imposition} for more details.

\subsection{Stress update algorithms}

The implementation of the two Bingham models requires prior knowledge of the strain-rate tensor, which can be approximated through the MPM discretization. At the start of each compute step, the strain-rate tensor at a material point can be calculated as
\begin{eqnarray}
    \tb D^{n+1}_p = \sym{\tb L_p^{n+1}}\,, \qquad \text{where} \qquad \tb L_p^{n+1} = \sum_{I=1}^{n_n}\nabla S_I(\tb x_p) \otimes \tb v^n_I\,.
    \label{eq:vel_grad}
\end{eqnarray}
Here, $\tb v^n_I$ denotes the nodal velocities interpolated from the particles during the P2G (particle-to-grid) mapping. The velocity-gradient tensor is also used to update the material point volume, according to:
\begin{eqnarray}
    V^{n+1}_p = \Delta J_p\,V_p^n\,, \qquad \text{where} \qquad \Delta J_p=1+\mathrm{tr}(\tb L_p^{n+1}) \Delta t\,,
\end{eqnarray}
and $\Delta t$ is the time-step increment, which must be sufficiently small to satisfy the Courant–Friedrichs–Lewy (CFL) condition for explicit time integration \hl{of the momentum balance equation}.

\subsubsection{Papanastasiou–Roussel Bingham fluid model}

Following the computation of the strain-rate tensor $\tb D^{n+1}$, the stress update for the PR-Bingham fluid model can be carried out explicitly at each material point following the classical hydrodynamics framework. For simplicity, the subscript $p$ is omitted hereafter, as all quantities are assumed to be material point properties. 

First, by evaluating the initial sound speed $c_0 = \sqrt{K_0 / \rho_0}$, the updated pressure $p^{n+1}$ can be computed using the weakly compressible equation of state previously given by Eq.~\eqref{eq: EOS1}, which governs the volumetric response of the material, i.e.,
\begin{equation}\label{eq: volumetric}
p^{n+1} = \frac{\rho_0 c_0^2}{\gamma}\left[\exp\left(-\gamma \varepsilon_v^{n+1} \right)-1\right]\,,
\qquad
\text{where}
\qquad
\varepsilon_v^{n+1}=\frac{V^{n+1}}{V_0}-1\,.
\end{equation}
Here, $V_0$ is the initial volume of the material point. \hl{Notice that for a weakly incompressible fluid, in line with the rigid-plastic assumption, the overall volumetric strain remains small, so that $J \approx 1$ and $\ln(J) \approx J-1$. Therefore, the first-order approximation considered in Eq.~\eqref{eq: volumetric} is an acceptable simplification.}

For the deviatoric response, the deviatoric component of the strain-rate tensor is used to compute the equivalent shear rate,
\begin{equation}
    \dot\gamma^{n+1} = \sqrt{2\boldsymbol{D}^{*,n+1}:\boldsymbol{D}^{*,n+1}}\,.
\end{equation}
This shear rate is then used to update the flocculation state parameter via a forward Euler time-integration scheme,
\begin{equation}
    \lambda^{n+1} = \lambda^n + \left(\frac{A_{\mathrm{thix}}}{\tau_0} - \alpha \lambda^n \dot\gamma^{n+1}\right)\Delta t\,.
    \label{eq:lambda_update}
\end{equation}
The initial value of the flocculation state $\lambda = \lambda_0$ should be assigned at the beginning of the time step, which can be determined either from rest time or direct experimental measurements. With the updated flocculation parameter, the deviatoric stress tensor $\tb \tau^{n+1}$ is then computed from the apparent viscosity and strain rates as:
\begin{equation}
   \boldsymbol{\tau}^{n+1} = 2\left[\eta + \frac{\tau_0}{\dot{\gamma}}(1+\lambda^{n+1}) (1 - e^{-m  \dot\gamma^{n+1} })\right]\boldsymbol{D}^{*,n+1}.
\end{equation}  

Finally, the total Cauchy stress is assembled by combining the deviatoric and isotropic components,
\begin{equation}\label{eq: total stress}
    \boldsymbol{\sigma}^{n+1} = \boldsymbol{\tau}^{n+1} - p^{n+1}\boldsymbol{I}.
\end{equation}

\begin{remark}
In the current study, we do not consider splitting the thixotropy evolution according to a critical shear rate, $\dot{\gamma}_c$, which is often employed to distinguish between the resting and flowing states of a material, such that regions with $\dot{\gamma} < \dot{\gamma}_c$ are treated as being at rest (e.g., \citet{beverly1992numerical}). Following this approach, as demonstrated by \citet{wilkes2023investigating}, thixotropic structuration is assumed to occur only in these resting states, whereas destructuration occurs only during the flowing states, i.e.~$\dot{\gamma} > \dot{\gamma}_c$. In contrast to their approach, the current model allows the thixotropy parameter to evolve, both structuration and destructuration, continuously, thereby eliminating the need for an explicit definition of a shear-rate threshold.
\end{remark}

\subsubsection{Elasto-viscoplastic Bingham model with thixotropy}

According to the theory established in Section \ref{sec:solid_model}, the stress-update algorithm of the elasto-viscoplastic model can be summarized as
\begin{equation}
    \tb \sigma^{n+1}= \tb \sigma^{n}  + \Delta t \Biggl(\mathbb{C}:\left(\tb D^{n+1} - \tb D^{p,n+1} \right) + \tb W^{n+1} \tb \sigma^{n} - \tb \sigma^{n} \tb W^{n+1}\Biggr)\,.
    \label{eq:total_stress_update}
\end{equation}
Here, the strain-rate tensor and the spin tensor are computed following Eq.~\eqref{eq:vel_grad}, with $\tb W^{n+1} = \skw{\tb L^{n+1}}$. The unknown in this expression is the plastic strain-rate tensor $\tb D^{p,n+1}$, whose value is zero for purely elastic deformation and nonzero otherwise, cf.~Eqs.~\eqref{eq:yield_cond} and \eqref{eq:flow_rule}.

The stress update procedure follows a predictor–corrector plastic flow formulation, in which a trial stress state $\tb{\sigma}^{tr}$ is first computed under the assumption of fully elastic deformation. If the trial stress lies within the \hl{static} yield surface \hl{(as determined by the first right-hand-side term on the right-hand side of Eq.~\eqref{eq:rate_dependent_yield_stress})}, it is accepted as the solution for the next step. Conversely, if the trial stress state exceeds the \hl{static} yield surface, a radial return-mapping algorithm is invoked based on the flow rule given in Eq.~\eqref{eq:flow_rule}. In addition, the evolution of the thixotropic state parameter $\lambda$ should be updated concurrently \hl{within the predictor-correction scheme}, which correspondingly causes the evolution of the effective yield stress $\tilde{\tau}_0$ with time and during plastic shearing. \hl{The flocculation state parameter is marched forward in time with a backward Euler integration scheme, consistent with the implicit return mapping algorithm, as
\begin{equation}
    \lambda^{n+1} = \lambda^n + \left(\frac{A_{\mathrm{thix}}}{\tau_0} - \alpha \lambda^{n+1} \dot\gamma^{p,n+1}\right)\Delta t\,.
    \label{eq:lambda_update_plastic}
\end{equation}
In contrast to the formulation assumed in the fluid model, Eq.~\eqref{eq:lambda_update}, we now employ the plastic strain rate $\dot{\gamma}^{p,n+1}$, which is initially unknown, instead of the total strain rate $\dot{\gamma}^{n+1}$, as previously introduced in Eq.~\eqref{eq:lambda_rate_plastic}.
}

The trial \hl{(or predictor)} stress $\tb \sigma^{tr}$ can be computed as:
\begin{eqnarray}
    \tb \sigma^{tr}= \tb \sigma^{n}  + \Delta t \Biggl(\mathbb{C}:\tb D^{n+1} + \tb W^{n+1} \tb \sigma^{n} - \tb \sigma^{n} \tb W^{n+1}\Biggr)\,.
\end{eqnarray}
Here, we consider a rotationally-neutralized stress update algorithm to ensure objectivity, as suggested by \citet{simo2006computational}. Considering that the motion is entirely elastic, we also compute the trial \hl{(or predictor)} thixotropy state parameter $\lambda^{tr}$ by setting $\dot{\gamma}^p=0$ \hl{in Eq.~\eqref{eq:lambda_update_plastic}}, i.e.,
\begin{equation}
    \lambda^{tr}=\lambda^n+\frac{A_{\mathrm{thix}}}{\tau_0}\Delta t\,.
\end{equation}
Here, only the flocculation rate is considered, as the structural build-up is a solely time-dependent function.

Once the trial stress and thixotropy state are obtained, the trial yield function $f^{tr}$ can be evaluated following Eq.~\eqref{eq:von_misses}, i.e.,
\begin{eqnarray}
    f^{tr}=\widehat{f}(\tb \sigma^{tr},\lambda^{tr}\hl{,\dot{\gamma}^p=0})=\tau^{tr} - (1+\lambda^{tr})\tau_0\,,
\end{eqnarray}
where $\tau^{tr}$ is the deviatoric stress invariant corresponding to $\tb \sigma^{tr}$. If $f^{tr} \leq 0$, the trial stress lies within the elastic domain and is therefore accepted as an admissible stress state. We then set $\tb \sigma^{n+1} = \tb \sigma^{tr}$, $\lambda^{n+1} = \lambda^{tr}$, and $\dot{\gamma}^{p,n+1} = 0$, accordingly.

\begin{figure}[h!]
    \centering
    \includegraphics[width=\linewidth]{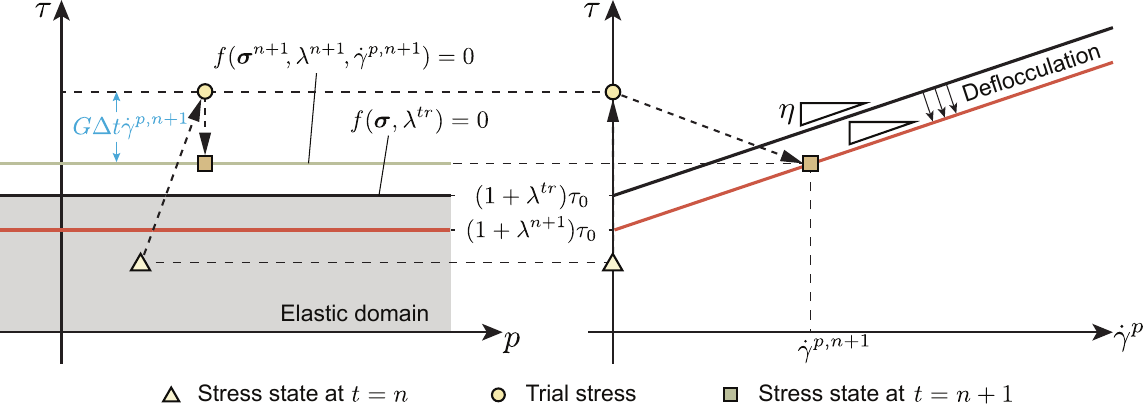}
    \caption{Schematic illustration of the proposed stress return mapping algorithm in $\tau$--$p$ space and $\tau$--$\dot{\gamma}^p$ space.}
    \label{fig:stress_return}
\end{figure}

Meanwhile, if $f^{tr} > 0$, the stress lies above the yield surface (see Fig.~\ref{fig:stress_return}), and hence, a return-mapping algorithm is required to bring the stress state back to the yield surface, i.e.~$\widehat{f}(\tb \sigma^{n+1}, \lambda^{n+1}, \dot{\gamma}^{p,n+1}) \to 0$. However, due to thixotropy evolution, the yield surface may soften according to the deflocculation term arising from plastic shearing (see the second term in Eq.~\eqref{eq:lambda_rate_plastic}). These conditions can be satisfied by minimizing the following set of corrector equations, \hl{which correspond to the remaining terms in Eqs.~\eqref{eq:total_stress_update} and \eqref{eq:lambda_update_plastic}}:
\begin{eqnarray}
\tb \sigma^{n+1} - \tb \sigma^{tr} + \Delta t \,\mathbb{C}:\tb D^{p, n+1}=\tb 0\,,
\label{eq:return_mapping_algebraic}\\
\lambda^{n+1}-\lambda^{tr}+ \Delta t\,\alpha \dot{\gamma}^{p,n+1}\lambda^{n+1}=0\,.
\label{eq:thixotropy_algebraic}
\end{eqnarray}
Here, note that $\dot{\gamma}^{p,n+1} > 0$, and consequently $\tb D^{p,n+1} \neq \tb 0$, as previously stated by the yield condition in Eq.~\eqref{eq:yield_cond} and the plastic flow rule in Eq.~\eqref{eq:flow_rule}.

As the adopted plastic flow rule is purely deviatoric (or isochoric), meaning no volumetric plastic deformation is considered, the updated pressure can be directly set as $p^{n+1} = p^{tr}$. Therefore, only the deviatoric stress needs to be computed through the return-mapping process. By taking the deviatoric invariant of Eq.~\eqref{eq:return_mapping_algebraic} and performing some algebraic manipulation, Eqs.~\eqref{eq:return_mapping_algebraic} and \eqref{eq:thixotropy_algebraic} can be reduced to a single residual equation:
\begin{eqnarray}
    \tau^{tr} - \left(\Delta t G + \eta\right) \dot{\gamma}^{p,n+1} - \tau_0\left(1 + \frac{\lambda^{tr}}{1+ \alpha \dot{\gamma}^{p,n+1} \Delta t}\right)=0\,.
    \label{eq:evp_residual}
\end{eqnarray}
Taking the positive root of $\dot{\gamma}^{p,n+1}$, we can obtain the closed-form expression:
\begin{eqnarray}
    \dot{\gamma}^{p,n+1} = \frac{
\sqrt{\xi^2 + 2\,\zeta\,f^{tr}} - \xi
}{
\zeta
}\,,
\label{eq:gamma_dot_thix}
\end{eqnarray}
where the constants $\xi$ and $\zeta$ are:
\begin{eqnarray}
    \xi = \eta + \Delta t \left(G + \alpha(\tau_0 - \tau^{tr})\right)\,, \qquad \zeta = 2\,\alpha\,\Delta t\,(\Delta t G + \eta)\,.
\end{eqnarray}

Once the plastic shear strain rate is obtained, the deviatoric stress invariant and the thixotropy state parameter can be updated as:
\begin{eqnarray}
    \tau^{n+1} = \tau^{tr} - \Delta t G \dot{\gamma}^{p,n+1}\,, \qquad \lambda^{n+1} = \frac{\lambda^{tr}}{(1+\Delta t \alpha \dot{\gamma}^{p,n+1})}\,.
\end{eqnarray}
Lastly, the total Cauchy stress can be assembled considering the co-directionality of the trial and updated deviatoric stresses, and thus,
\begin{equation}
    \boldsymbol{\sigma}^{n+1} = \frac{\tau^{n+1}}{\tau^{tr}}\boldsymbol{\tau}^{tr} - p^{n+1}\boldsymbol{I}.
\end{equation}

\begin{remark}
    It is worth noting that Eq.~\eqref{eq:gamma_dot_thix} is derived under the assumption that $\lambda$ varies with the plastic shear strain rate $\dot{\gamma}^p$, i.e.~when $\alpha >0$. When $\alpha = 0$, corresponding to a non-thixotropic material, the denominator $\zeta = 0$, resulting in Eq.~\eqref{eq:gamma_dot_thix} being undefined. In this case, the correct procedure is to set $\alpha = 0$ directly in Eq.~\eqref{eq:evp_residual}. This reduces the residual to a linear equation, leading to the explicit solution if $\alpha=0$:
\begin{eqnarray}
    \dot{\gamma}^{p,n+1} = \frac{f^{tr}}{\eta+\Delta t G}\,.
\label{eq:gamma_dot_no_thix}
\end{eqnarray}
\end{remark}
\begin{remark}
    \hl{For each time step, the computational cost of the elasto-viscoplastic model is approximately the same as that of the PR-Bingham fluid model, as both require only a local, closed-form stress update at each material point. However, the two models differ in time-step constraints. For the fluid model, the apparent viscosity approaches $\eta_{a} \to \eta + m\tau_0$ in the limit of $\dot\gamma \to 0$ (cf.~Eq.~\eqref{eq:regularized_bingham}), so a larger value of $m$, which may be needed for better accuracy, proportionally tightens the critical time step as $\Delta t_{\mathrm{crit}} \propto 1/\eta_a \propto 1/m$. In contrast, the elasto-viscoplastic model avoids this trade-off, since its stability is governed solely by the physical elastic wave speed, and thus does not depend on any regularization parameter that influences computational stability or cost.}
\end{remark}




\subsection{Verification test: planar Poiseuille flow}
\label{subsec:poiseuille_flow}

\begin{figure}[h!]
    \centering
    \includegraphics[width=\linewidth]{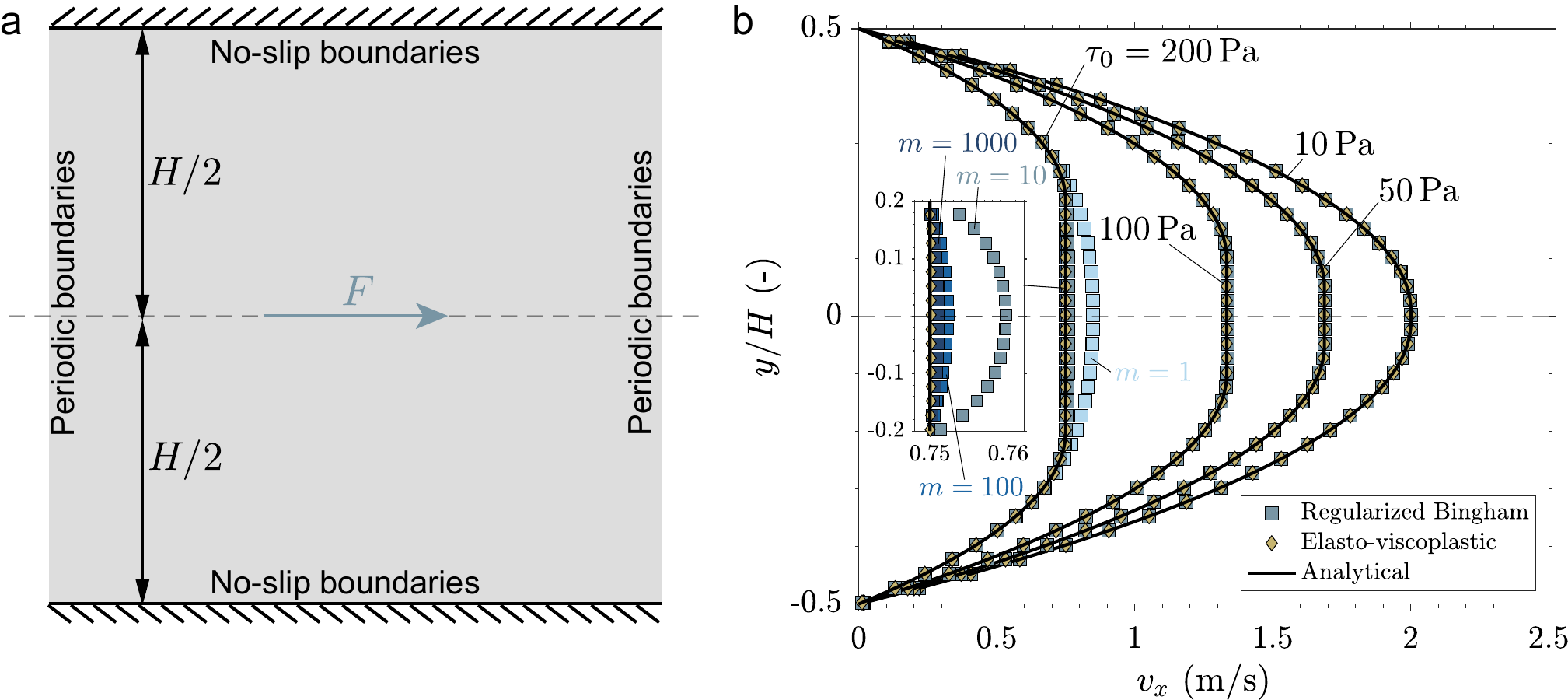}
    \caption{Planar Poiseuille flow: (a) Model geometry and (b) obtained numerical results utilizing the Papanastasiou-regularized and elasto-viscoplastic Bingham model in comparison with analytical solutions. The inset plot shows a zoomed view of the plug velocity field with $\tau_0=200$ Pa. \hl{The sensitivity analysis for various regularization parameters $m$ at $\tau_0 = 200$ Pa is represented by square markers in different shades of blue.}}
    \label{fig:poiseuille}
\end{figure}

To verify the proposed formulation and implementation, we conducted a one-dimensional Poiseuille flow simulation between parallel plates. The geometry of the problem is shown in Fig.~\ref{fig:poiseuille}a, where a Bingham fluid is confined between two parallel plates with no-slip boundaries. A horizontal body force per unit volume, $F$, is applied to drive the flow until a steady-state condition is reached. To obtain a unique steady-state solution, the thixotropic model is disabled by keeping the thixotropy parameter unchanged, that is $\lambda=0$ and $\pd\lambda/\pd t = 0$ throughout the simulation. The steady-state horizontal velocity solution is given for $y\in(-H/2,\,H/2)$ as \citep{bird2002transport}:
\begin{eqnarray}
v_x=
\begin{cases}
\hat{v}_x(y_0), & |y|\le y_0\,, \\
\hat{v}_x(|y|), & y_0 < |y| \le \dfrac{H}{2}\,,
\end{cases}
\end{eqnarray}
where the boundary between the yielding and non-yielding regions is defined by $y_0 ={\tau_0}/{F}$. The velocity profile of the yielded region is
\begin{eqnarray}
\begin{split}
\hat{v}_x(|y|)
&=
\frac{1}{2\eta}
\left[
F\!\left(\frac{H^2}{4} - y^2\right)
- 2\tau_0 \left(\frac{H}{2}-|y| \right)
\right]\,.
\end{split}
\end{eqnarray}
Meanwhile, the \textit{plug-region} velocity is constant and equal to:
\begin{eqnarray}
\begin{split}
\hat{v}_x(y_0)
&=
\frac{1}{2\eta}
\left[
F\!\left(\frac{H^2}{4}-y_0^2\right)
- 2\tau_0\left(\frac{H}{2} - y_0\right)
\right]\,.
\end{split}
\end{eqnarray}

The material and loading parameters used in this verification test are: the yield stress is considered to vary $\tau_0=\{10,\,50,\,100,\,200\}$ Pa, mass density $\rho=2000$ kg/m$^3$, body force $F=2000$ N/m$^3$, channel height $H=0.5$ m, and dynamic viscosity $\eta=30$ Pa$\cdot$s. For the Papanastasiou Bingham model, the regularization parameter is set to $m=10$ by default. \hl{An accompanying sensitivity analysis is also performed by varying $m=\{1,\,10,\,100,\,1000\}$ for the case with $\tau_0=200$ Pa.} The initial bulk modulus is $K_0=0.33$ MPa, and the Poisson's ratio is $\nu=0.45$. The simulation is performed using MPM with periodic boundary conditions along the sides, 16 particles per cell, and a cell size of 0.01 m. The scheme and other numerical settings follow those described in Section \ref{sec: Numerical examples}. The simulation results from both the Papanastasiou Bingham model and the elasto-viscoplastic model are shown in Fig.~\ref{fig:poiseuille}b, where the characteristic plug flow at the channel center is accurately reproduced. \hl{In this study, we aim to verify and show that the proposed framework and its implementation are consistent with the Bingham fluid model in the steady-state flow regime, and that they achieve even higher accuracy than the regularized Bingham model, even when using a large value of the regularization parameter $m$ in capturing the plug velocity field, for which the shear strain rate should be exactly zero.}

\section{Numerical examples}
\label{sec: Numerical examples}
In this section, we rigorously validate the proposed model against several common on-site testing procedures for fresh concrete. We first conducted a series of slump flow tests \citep{BS123508}  with three primary objectives: (1) to evaluate whether the proposed elasto-viscoplastic model, stress-integration algorithm, and MPM implementation can accurately capture the flow–rest transition of fresh concrete and the resulting slump flow distance; (2) to assess the capability of the model in reproducing the dynamic flow process; and (3) to verify whether the proposed thixotropic model can consistently simulate the thixotropic behavior of concrete, including the rheological response after different rest times. Finally, we simulated two additional standard tests, the L-box test and the V-funnel test, to further assess the model’s capability in predicting flow interacting with reinforcing bars and in capturing flow under channelized geometries.

In all MPM simulations presented in this section, we employ the explicit MPM formulation with linear shape functions and three-dimensional structured hexahedral elements. To ensure numerical stability and mitigate volumetric locking under nearly incompressible conditions, we adopt the MUSL (Modified Update Stress Last) scheme \citep{nairn2003material} together with the B-bar method \citep{hughes1980generalization}. For particle velocity updates, the FLIP (FLuid Implicit Particle) scheme \citep{brackbill1986flip} is used to maintain energy conservation. \hl{In addition, the regularization parameter of the PR-Bingham model is fixed at $m=10$ throughout the numerical examples presented in this section, unless otherwise stated \citep{wilkes2023investigating}.} The current MPM implementation and framework are available as open source on GitHub\footnote{https://github.com/geomechanics/mpm}, and have been extensively developed and validated for a wide range of solid, geomechanics, and fluid mechanics applications \citep{kularathna2021semi, liang2022shear, chandra2024mixed, chandra2024stabilized, kurima2025absorbing}.


\subsection{Slump flow test}
The slump flow test is widely adopted as an on-site method to assess the workability and flow characteristics of fresh concrete, particularly self-compacting and Tremie concrete mixes \citep{domone1998slump, roussel2016, wilkes2023investigating}. It provides a practical and straightforward field test to quantify concrete flowability by measuring the final slump flow diameter (denoted as SF) after the standard cone is lifted. For Tremie concrete, which is essential in deep foundation applications such as diaphragm walls and bored piles, maintaining high fluidity is crucial to ensure proper flow around reinforcement cages and through pipes without segregation. In this study, three sets of slump flow simulations are conducted, each designed with a specific objective.

\subsubsection{Roussel's benchmark test}
\label{subsubsec:roussel_benchmark}
\begin{figure}[h!]
    \centering
    \includegraphics[width=1\linewidth]{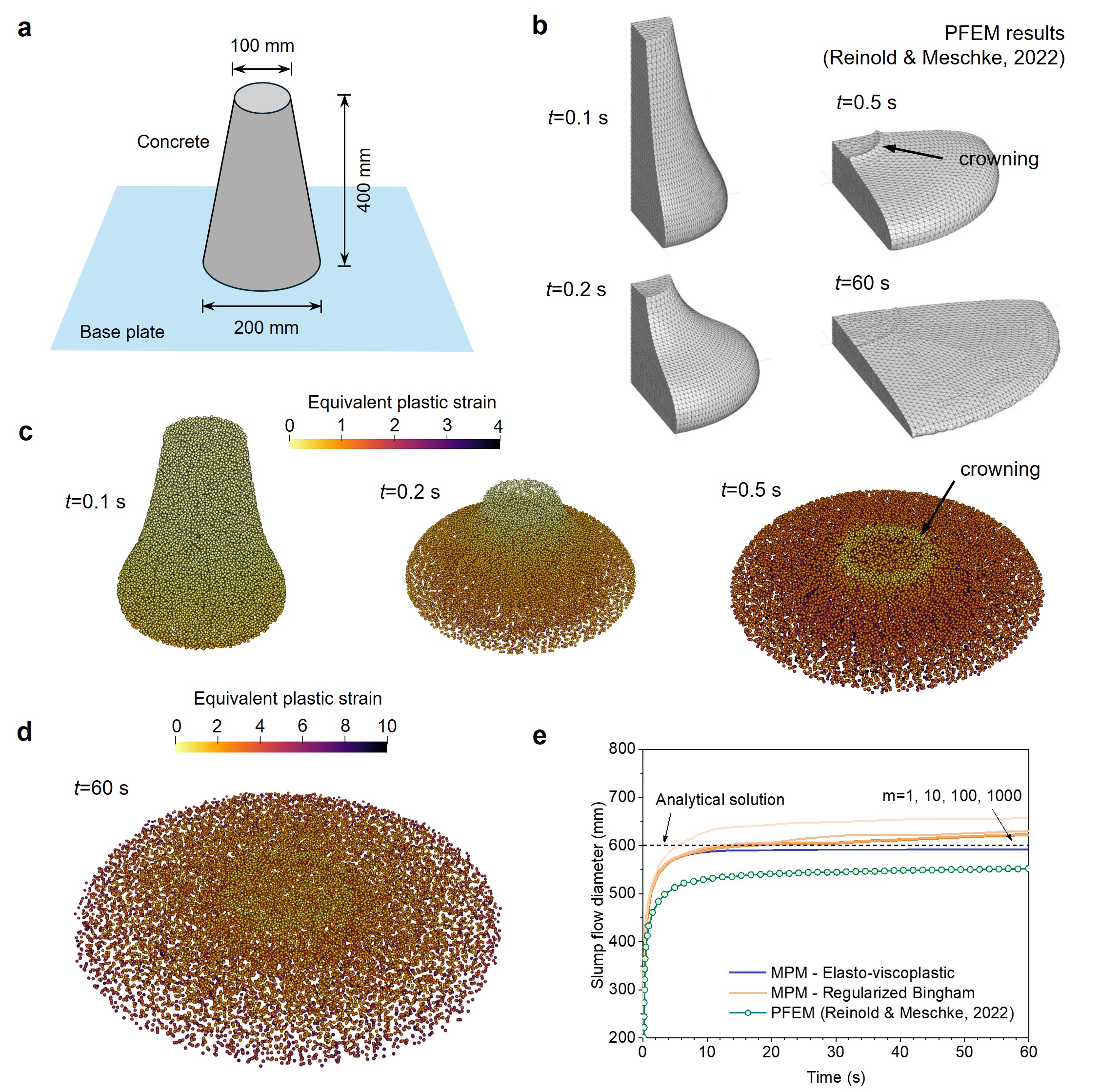}
    \caption{Roussel's slump flow benchmark \citep{roussel2016}: (a) Geometry of the concrete cone; (b) simulation results of the concrete slump flow by PFEM \citep{reinold2022}; (c) simulation results of the concrete slump flow at $t =$ 0.1, 0.2, and 0.6 s; (d) final shape of the concrete flow ($t =$ 60 s); and (e) comparion of the evolution of the slump flow diameter using the elasto-viscoplastic solid model and the regularized Bingham fluid model \hl{with different regularization parameter $m$}. The crowning-like region highlighted in (b) and (c) shows the unsheared (plug) domain at the top tip of the concrete slump.}
    \label{fig: Roussel_benchmark-1}
\end{figure}

Roussel's slump flow test is a well-known benchmark for evaluating numerical models of fresh concrete flow \citep{roussel2016} across different numerical simulators. The test is based on the standard Abrams cone geometry, with initial dimensions $H_0 = 300\, \text{mm}$, $R_{\text{min}} = 50 \, \text{mm}$, and $R_{\text{max}} = 100 \, \text{mm}$, as shown in Fig.~\ref{fig: Roussel_benchmark-1}a. The fresh concrete is modeled as a homogeneous Bingham material characterized by a yield stress $\tau_0 = 50 \, \text{Pa}$, plastic viscosity $\eta = 50 \, \text{Pa·s}$, mass density $\rho = 2300 \, \text{kg/m}^3$, and gravitational acceleration $g = 9.81 \, \text{m/s}^2$. The analytical solution for the final slump flow diameter is derived under the following assumptions \citep{roussel2005fifty}: (1) inertia and surface tension effects are neglected; (2) the flow regime is quasi-static; and (3) the problem is simplified to a one-dimensional flow profile in the regime $H \ll R$ with a one-dimensional yielding criterion where flow ceases once the shear stress equals the yield stress. Here, $H$ and $R$ are the final height and radius of the specimen. The final surface profile is expressed as:
\begin{equation}
    h(r) = \frac{2{\tau_0}}{\rho g} (R - r)\,,\qquad 
    {R = \left(\frac{225 \rho g \Omega^2}{128 \pi^2 \tau_0}\right)^{1/5}}\,,
\end{equation}
where $h(r)$ is the height or thickness of the concrete at radial distance $r$ and $\Omega$ is the specimen volume (can be taken as its initial value). The effect of thixotropy is not considered in this benchmark test.

\begin{figure}[h!]
    \centering
    \includegraphics[width=1\linewidth]{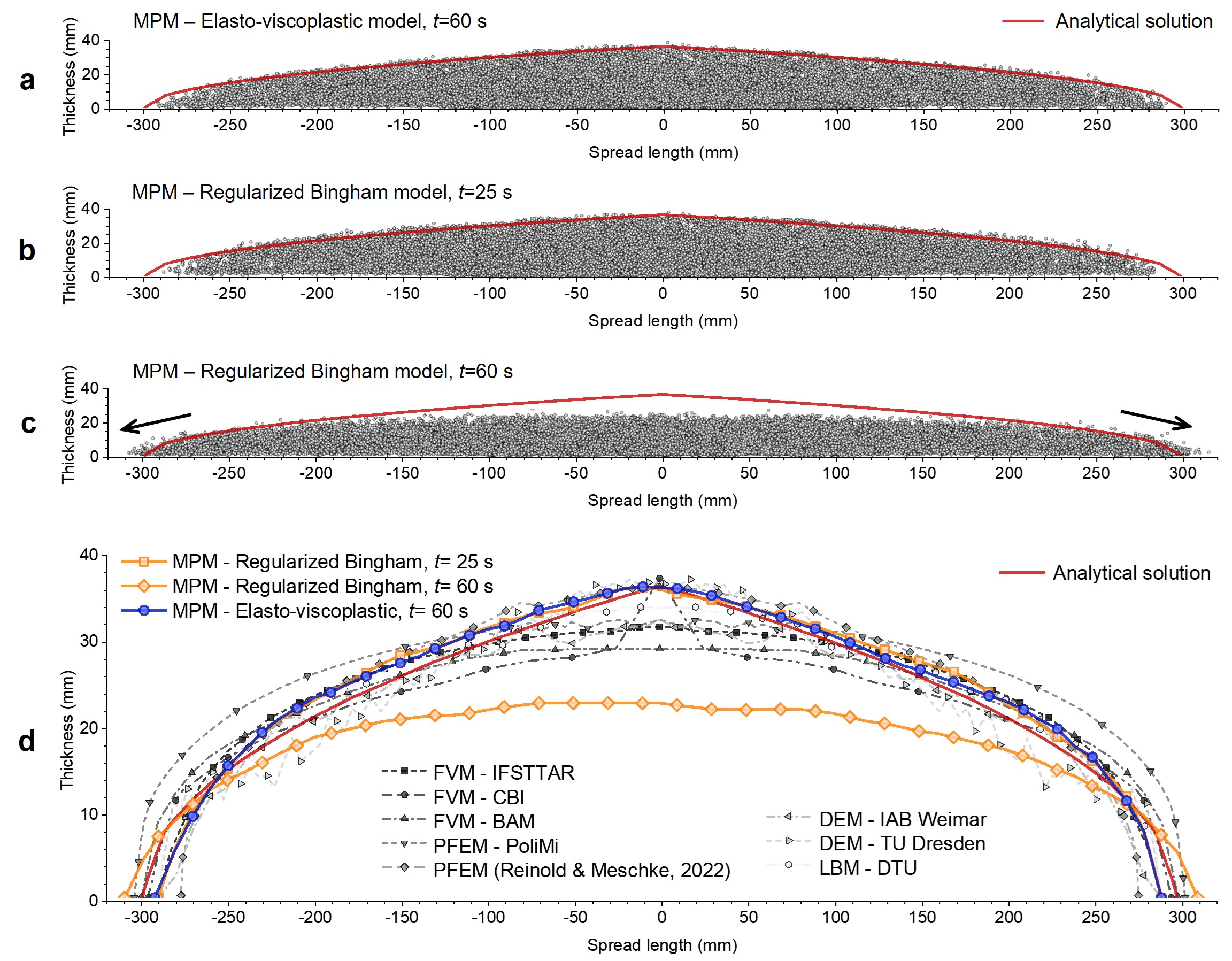}
    \caption{Roussel's slump flow benchmark \citep{roussel2016}: Comparison between MPM simulation results and analytical solution for surface profile: (a) elasto-viscoplastic model at $t =$ 60 s, (b) regularized Bingham model at $t =$ 25 s, and (c) regularized Bingham model at $t =$ 60 s. (d) Comparison of MPM simulation results with other numerical methods. Black arrows in (c) highlight that the concrete slump continues to flow outward even at $t=60$ s.}
    \label{fig: Roussel_benchmark-2}
\end{figure}

In the MPM simulation, a no-slip boundary condition is applied at the flat base, while the influence of the cylindrical container surrounding the concrete is neglected \citep{reinold2022}. The initial material points are generated using an unstructured distribution, where tetrahedral elements are first created with {\texttt{Gmsh} \citep{geuzaine2009gmsh}}. A material point is then placed at the centroid of each tetrahedron, and its volume is assigned to match that of the corresponding element. The simulation contains a total of 24,027 material points, and the background mesh consists of a structured hexahedral grid with a cell size of 10 mm. The elastic parameters are taken as Young's modulus {$E_0 = 100$} kPa and Poisson's ratio $\nu = 0.3$ following \cite{reinold2022}\footnote{Knowing Young's modulus and the Poisson's ratio, one may compute the initial bulk modulus $K_0$ required for the volumetric response (cf. Eq.~\eqref{eq: EOS1}).}. The time step is set to $5 \times 10^{-4}$ s. \hl{For the PR-Bingham fluid model, we tested a range of regularization parameters $m$, from 1 to 1000. The results show that the slump flow distance generally increases with decreasing $m$ (Fig.~\ref{fig: Roussel_benchmark-1}e), indicating that the simulation results of the regularized Bingham model are sensitive to the choice of the regularization parameter. In the following analysis, we therefore adopt $m=10$ as the reference case, consistent with the value suggested in the literature \citep{wilkes2023investigating}.}

Fig.~\ref{fig: Roussel_benchmark-1}c presents the simulation results of the slump flow test at different time instants using the elasto-viscoplastic model, while Fig.~\ref{fig: Roussel_benchmark-1}d shows the final concrete profile, with the colored contour representing the equivalent plastic strain. For comparison, the simulation results by PFEM reported in \cite{reinold2022} are also presented in Fig.~\ref{fig: Roussel_benchmark-1}b (only a quarter of the domain is simulated). It can be observed that the slump flow shapes predicted by both MPM and PFEM are quite similar. Both our simulation and the PFEM simulation exhibit a crowning-like region around $t = 0.5$ s, indicating a circular plug zone undergoing rigid-body motion with no plastic shear strain. In Fig.~\ref{fig: Roussel_benchmark-1}e, a further comparison is made between the simulated final slump flow diameter and the analytical solution. The results indicate that the MPM simulation based on the elasto-viscoplastic model stabilizes after 10 s, and the final value agrees very well with the analytical solution. In contrast, the MPM simulation using the regularized Bingham fluid model shows a continuous increase in the slump flow diameter, eventually exceeding the analytical prediction. Additionally, the PFEM results based on a hyperelastic-viscoplastic model can also capture the stoppage of the flow, though the simulated values appear to be smaller than the analytical solution and our simulation results. This discrepancy may be attributed to numerical diffusion associated with the smoothing and projection of internal variables during re-meshing in their PFEM framework, as discussed by the authors themselves.

In Fig.~\ref{fig: Roussel_benchmark-2}, a more detailed comparison of the cross-section profile between the simulations and the analytical solution is provided. It can be observed that the MPM simulation using the elasto-viscoplastic model matches the analytical solution with good accuracy (Fig.~\ref{fig: Roussel_benchmark-2}a). The simulation using the regularized Bingham fluid model also shows good agreement at $t = 25$ s (Fig.~\ref{fig: Roussel_benchmark-2}b). However, because the fluid model continues to flow to carry shear stress, the simulation at $t = 60$ s exhibits noticeable deviation and an overestimation of the spread (Fig.~\ref{fig: Roussel_benchmark-2}c). This spreading trend would continue if the simulation were extended further. 

In Fig.~\ref{fig: Roussel_benchmark-2}d, results from various methods compiled by \citet{roussel2016} are compared. Although most of these methods are based on fluid models and provide reasonably good predictions, they often rely on artificial flow-stop criteria (for example, terminating the simulation at 20 s) \citep{roussel2016, wilkes2023investigating}, chosen solely for convenience and without physical justification. Such approaches, however, are unsuitable for long-duration simulations, including real-time casting scenarios. In contrast, the proposed elasto-viscoplastic model accurately captures the cessation of concrete flow and the final spread without requiring any ad hoc stopping criteria.

\subsubsection{Validation of dynamic evolution of slump flow}
\label{sec: Hao's experiment}
\begin{figure}[h!]
    \centering
    \includegraphics[width=1\linewidth]{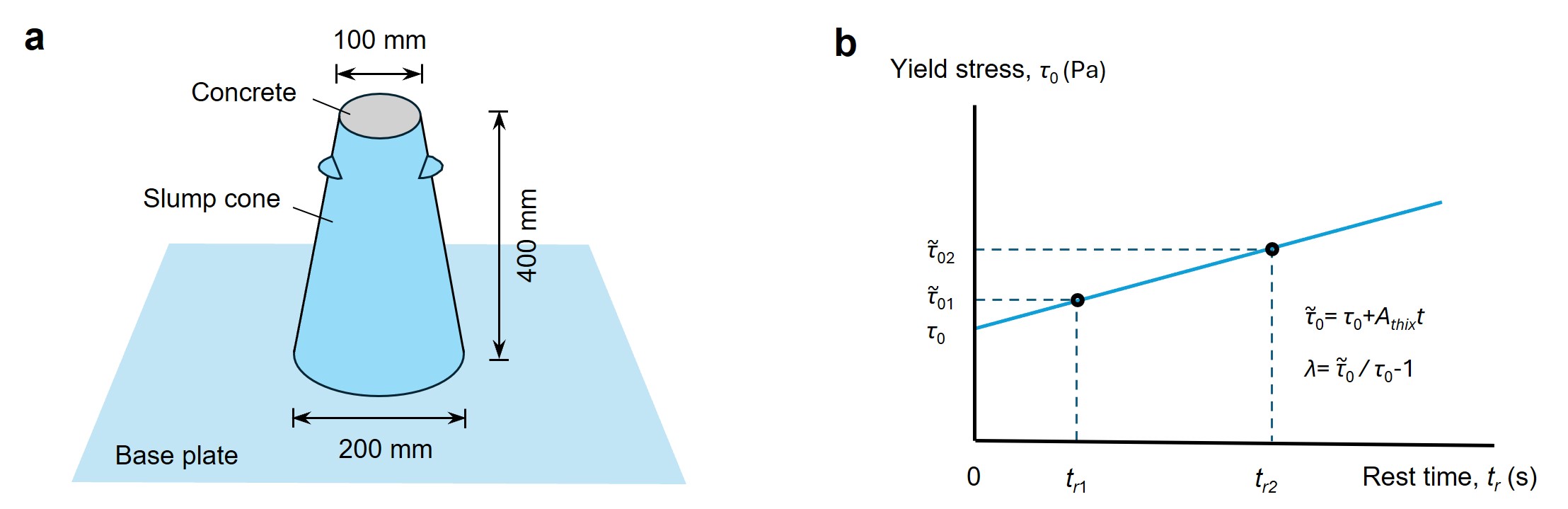}
    \caption{Dynamic evolution of slump flow: (a) Model setup of the slump flow test in the experiment. (b) Linear fitting of the thixotropic parameters. Effective yield stresses $\tilde{\tau}_{01}$ and $\tilde{\tau}_{02}$ are measured at $t_{r1}$ and $t_{r2}$, and $\tau_0$, $A_{\mathrm{thix}}$, and initial $\lambda$ values are derived.}
    \label{fig: Haos_experiment-1}
\end{figure}

In Roussel’s benchmark test discussed in Section \ref{subsubsec:roussel_benchmark}, the analytical solution is derived under quasi-static assumptions and therefore only provides the final surface profile, offering no means of validating the dynamic evolution of the flow. However, accurately capturing the transient flow behavior is crucial for simulating concrete casting processes. To evaluate the capability of our model in this regard, we consider and simulate a series of laboratory slump flow tests conducted by \cite{luo2014cement}. In these laboratory tests, an Abrams cone is placed on a flat surface, filled with fresh \hl{cement paste} as illustrated in Fig.~\ref{fig: Haos_experiment-1}a, and then lifted vertically at a controlled speed to allow the material to flow outward in all directions. The entire flow process is recorded using high-speed cameras, and the measured SF distances are compared with the numerical simulations.

\begin{table}[h!]
    \caption{Dynamic evolution of slump flow: Summary of test conditions, material parameters, and measured slump flow diameter in experiments.}
    \centering
    {
    \footnotesize
    \begin{tabular*}{\textwidth}{@{\extracolsep{\fill}}cccccccccccc@{}}
        \toprule
        \multicolumn{2}{c}{\hl{Cement mix ID}} & Test & $t_r$ & $\rho$ & {$\tilde{\tau}_0$} & {$\tau_{0}$} & $A_{\mathrm{thix}}$ & $\lambda$ & $\eta$ & $\mu_f$ & $\text{SF}_{\text{exp}}$ \\ 
        \multicolumn{2}{c}{} & & (s) & $(\text{kg/m}^3)$ & (Pa) & (Pa) & (Pa/s) & - & (Pa s) & - & (mm)\\ 
        \midrule
        \multirow{2}{*}{Mix-1} & CEMI: & SF30 & 30 & \multirow{2}{*}{1820} & 74.9 & \multirow{2}{*}{16.8} & \multirow{2}{*}{1.94} & 3.45 & \multirow{2}{*}{25*} & \multirow{2}{*}{0.3} & 529 \\
        & WC0.5-Bt4\% & SF120 & 120 &  & 249.1 &  &  & 13.8 &  &  & 416 \\
        \midrule
        \multirow{2}{*}{Mix-2} & CEMII: & SF30 & 30 & \multirow{2}{*}{1820} & 101.5 & \multirow{2}{*}{67.8} & \multirow{2}{*}{1.12} & 0.50 & \multirow{2}{*}{23{*}} & \multirow{2}{*}{0.3} & 497.5 \\
        & WC0.5-Bt4\% & SF120 & 120 &  & 202.6 &  &  & 1.99 &  &  & 433 \\
        \midrule
        \multirow{2}{*}{Mix-3} & CEMII: & SF30 & 30 & \multirow{2}{*}{1440} & 6.9 & \multirow{2}{*}{2.97} & \multirow{2}{*}{0.13} & 1.33 & \multirow{2}{*}{10{*}} & \multirow{2}{*}{0.3} & 810 \\
        & WC1.0-Bt8\% & SF120 & 120 &  & 18.7 &  &  & 5.30 &  &  & 665 \\
        \midrule
        \multirow{2}{*}{Mix-4} & CEMII: & SF30 & 30 & \multirow{2}{*}{1830} & 1000 & \multirow{2}{*}{900} & \multirow{2}{*}{-} & - & \multirow{2}{*}{50{*}} & \multirow{2}{*}{0.3} & 244 \\
        & WC0.5-Bt8\% & SF120 & 120 &  & 1300 &  &  & - &  &  & 219 \\
        \bottomrule
    \end{tabular*}
    }
    \begin{tablenotes}
    \item[*] \footnotesize {* Calibrated value.}
    \item[*] \footnotesize Note: WC - water-to-cement ratio; Bt - bentonite content; SF30/SF120 - slump flow with rest time of 30 s/120 s; $t_r$: - rest time; $\rho$ - density of concrete mix; {$\tilde{\tau}_0$ - effective yield stress at $t_r>0$}; {${\tau}_0$ - initial yield stress at $t_r=0$}; $A_{\mathrm{thix}}$ - flocculation rate; $\lambda$ - flocculation state variable; $\eta$ - plastic viscosity; $\mu_f$ - friction coefficient of base plate; and $\text{SF}_{\text{exp}}$ - experimental slump flow diameter.
    \end{tablenotes}
\label{table: 1}
\end{table}

We selected four slump flow tests as validation cases, as summarized in Table \ref{table: 1}. These cases include two types of cement (i.e.~CEM I and CEM II), two water-to-cement ratios (i.e.~W/C = 0.5 and 1.0), and two bentonite contents (i.e.~Bt = 4\% and 8\%). Generally, a higher water-to-cement ratio increases the flowability of cement paste, whereas higher bentonite content decreases it. For each mix, slump flow tests are conducted at two rest times, $t_r = 30$ and $120 \, \text{s}$, denoted as SF30 and SF120, respectively. Based on the {effective yield stress ($\tilde{\tau}_0$)} obtained from SF30 and SF120, the {initial yield stress ($\tau_0$)} and the flocculation rate ($A_{\text{thix}}$) are estimated through linear fitting (see Fig.~\ref{fig: Haos_experiment-1}b). {For all the tests in this section, the deflocculation rate, $\alpha$, is set to a small value of 0.01 \citep{wilkes2023investigating}.} The initial flocculation state variable $\lambda$ can be determined accordingly. \hl{It is important to note that, due to the lack of relevant experimental data, the plastic viscosity ($\eta$) is calibrated for each cement mix, while the base plate friction coefficient ($\mu_f$) is fixed at a representative value of 0.3 for all cases.} Furthermore, to accurately replicate the slump flow process observed in the laboratory, the bucket outside the concrete cone (mold) is modeled using rigid particles, and the lifting speed is controlled at {0.15~m/s}. In addition, the elastic parameters of the mixes are set as follows: Young’s modulus is set to $E = 1\times10^5$ Pa and Poisson’s ratio to $\nu = 0.45$, \hl{values that are widely used in the literature for fresh concrete or cementitious materials \citep{roussel2012, bolton1997geotechnical, wilkes2023investigating}}. A time step of $\Delta t = 5\times10^{-4}$ s is used in the simulation.

\begin{figure}[h!]
    \centering
    \includegraphics[width=1\linewidth]{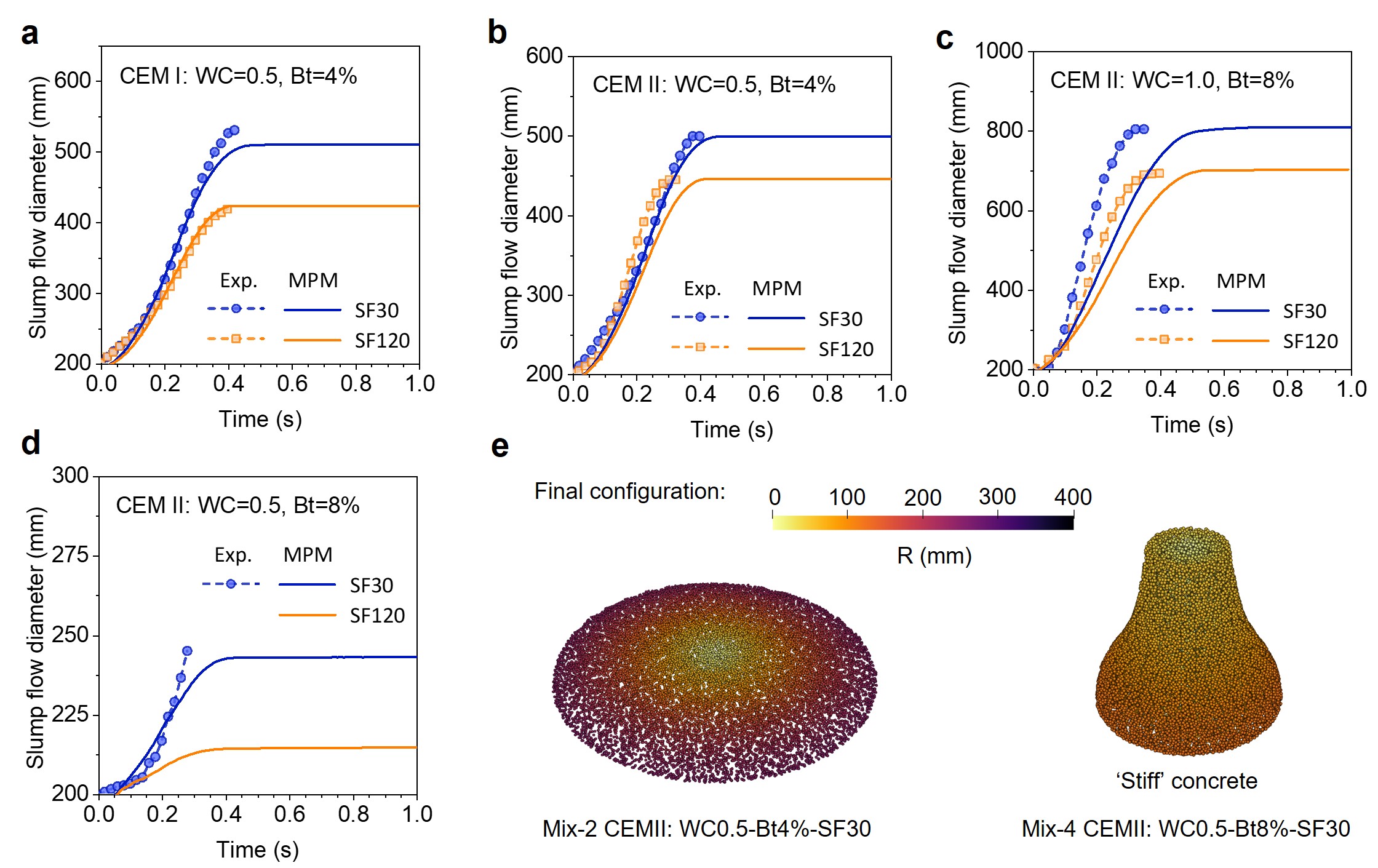}
    \caption{Dynamic evolution of slump flow: (a-d) Comparison of the dynamic evolution of slump flow diameter obtained by MPM simulation and laboratory test. (e) Final simulation snapshot from Mix-2 and 4.}
    \label{fig: Haos_experiment-2}
\end{figure}

\begin{figure}[h!]
    \centering
    \includegraphics[width=1\linewidth]{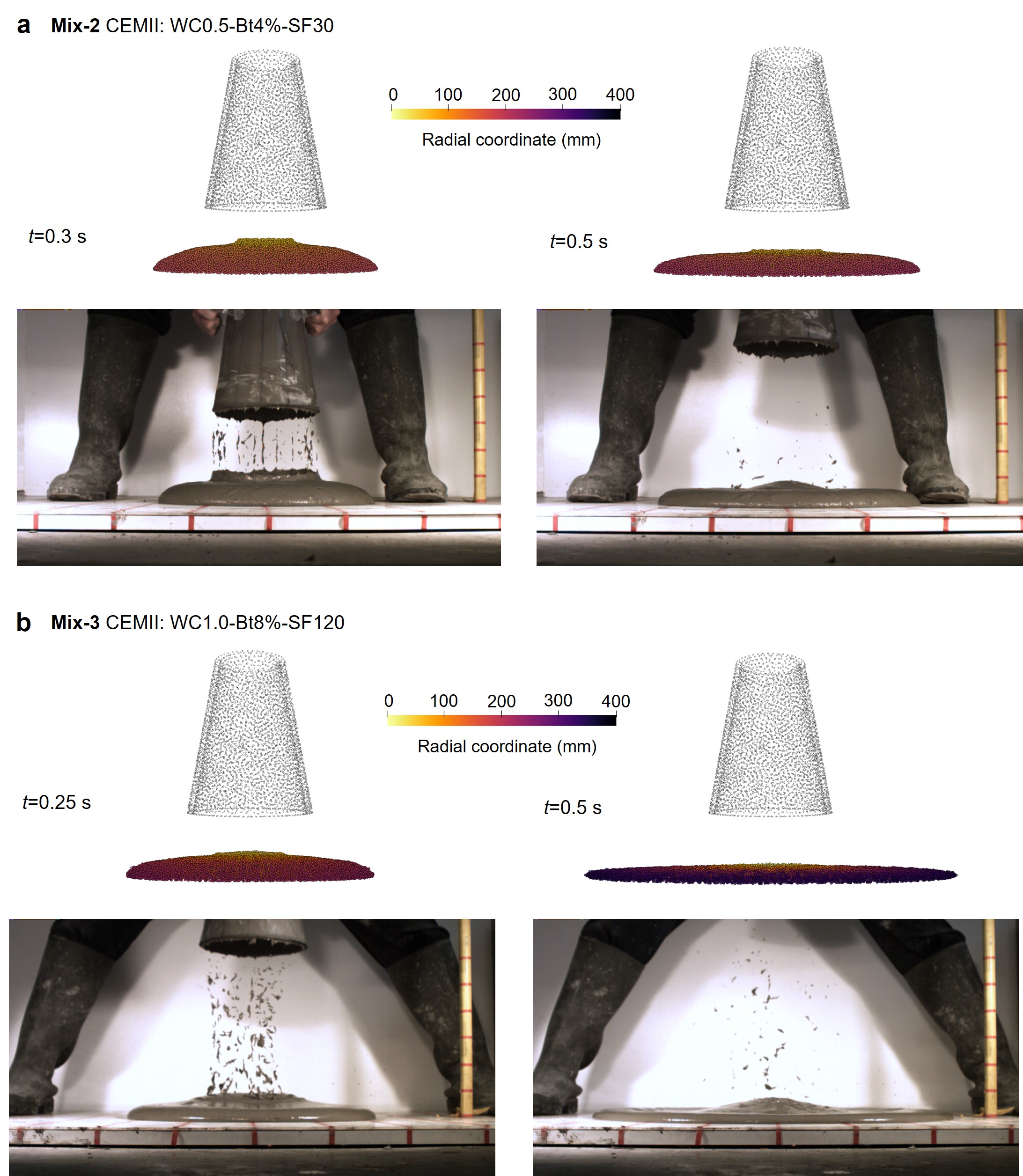}
    \caption{Dynamic evolution of slump flow: Comparison of the slump profiles from MPM simulations and laboratory tests at $t = 0.3$ s and $t = 0.5$ s for (a) Mix-2 (CEMII: WC0.5-Bt4\%-SF30) and (b) Mix-3 (CEMII: WC1.0-Bt8\%-SF120). Experiment photos are taken from \cite{luo2014cement}.}
    \label{fig: Haos_experiment-3}
\end{figure}

Fig.~\ref{fig: Haos_experiment-2}a-d compares the obtained simulation results and experimental data for the four mixes. From the simulation results, it is observed that for all cases, the slump flow diameter of SF120 is smaller than that of SF30. This result is expected, as the additional 90 seconds of rest time increases the yield stress of the SF120 case, thereby reducing its flowability. Furthermore, the evolution of the SF diameter initially increases slowly, likely due to the constraint imposed by the cone boundary, then accelerates, and finally decelerates, with all cases ceasing flow after approximately 0.4 s. This demonstrates the capability of the elasto-viscoplastic model to effectively capture the cessation of slump flow. For Mix-1 and Mix-2 (see Fig.~\ref{fig: Haos_experiment-2}a and b), the simulated SF diameter evolution agrees well with experimental measurements, with final diameters in the range of 400–600 mm, which is typical for tremie concrete slump flow. This demonstrates that our model can not only accurately predict the final slump flow diameter but also capture the entire flow dynamics. In the case of Mix-3 and Mix-4 (see Fig.~\ref{fig: Haos_experiment-2}c and d), which exhibit significantly stronger and weaker flowability, respectively, the simulation shows slight deviations from experimental data; however, the overall trends remain consistent, and the discrepancies are within acceptable ranges.

\begin{figure}[h!]
    \centering
    \includegraphics[width=1\linewidth]{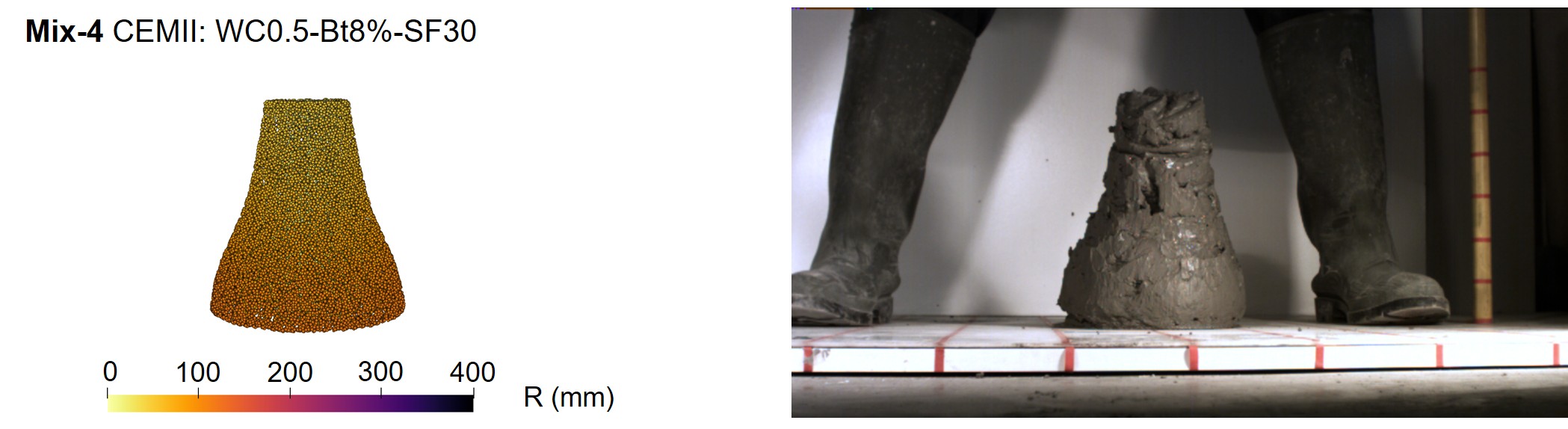}
    \caption{Dynamic evolution of slump flow: Comparison of the slump profiles from MPM simulation and laboratory test at $t = 1.0$ s for Mix-4 (CEMII: WC0.5-Bt8\%-SF30), a less workable cement paste. Experiment photos are taken from \cite{luo2014cement}.}
    \label{fig: Haos_experiment-4}
\end{figure}

Fig.~\ref{fig: Haos_experiment-2}e presents the final simulated profiles of a more workable \hl{cement paste} (Mix-1) and a less workable \hl{cement paste} (Mix-4), illustrating the model’s ability to replicate the distinct flow behaviors associated with different levels of workability. Additionally, as shown in Figs.~\ref{fig: Haos_experiment-3}–\ref{fig: Haos_experiment-4}, comparisons between the simulated profiles and photographed experimental results for Mix-2-SF30, Mix-3-SF120, and Mix-4-SF30 further validate the model’s accuracy across a range of mixtures, from highly flowable to low-flowability compositions. These results demonstrate the versatility and robustness of the proposed modeling framework.

\subsubsection{Modeling the thixotropic behavior by slump flow tests}
\label{sec: slump flow test}
\begin{figure}[h!]
    \centering
    \includegraphics[width=0.55\linewidth]{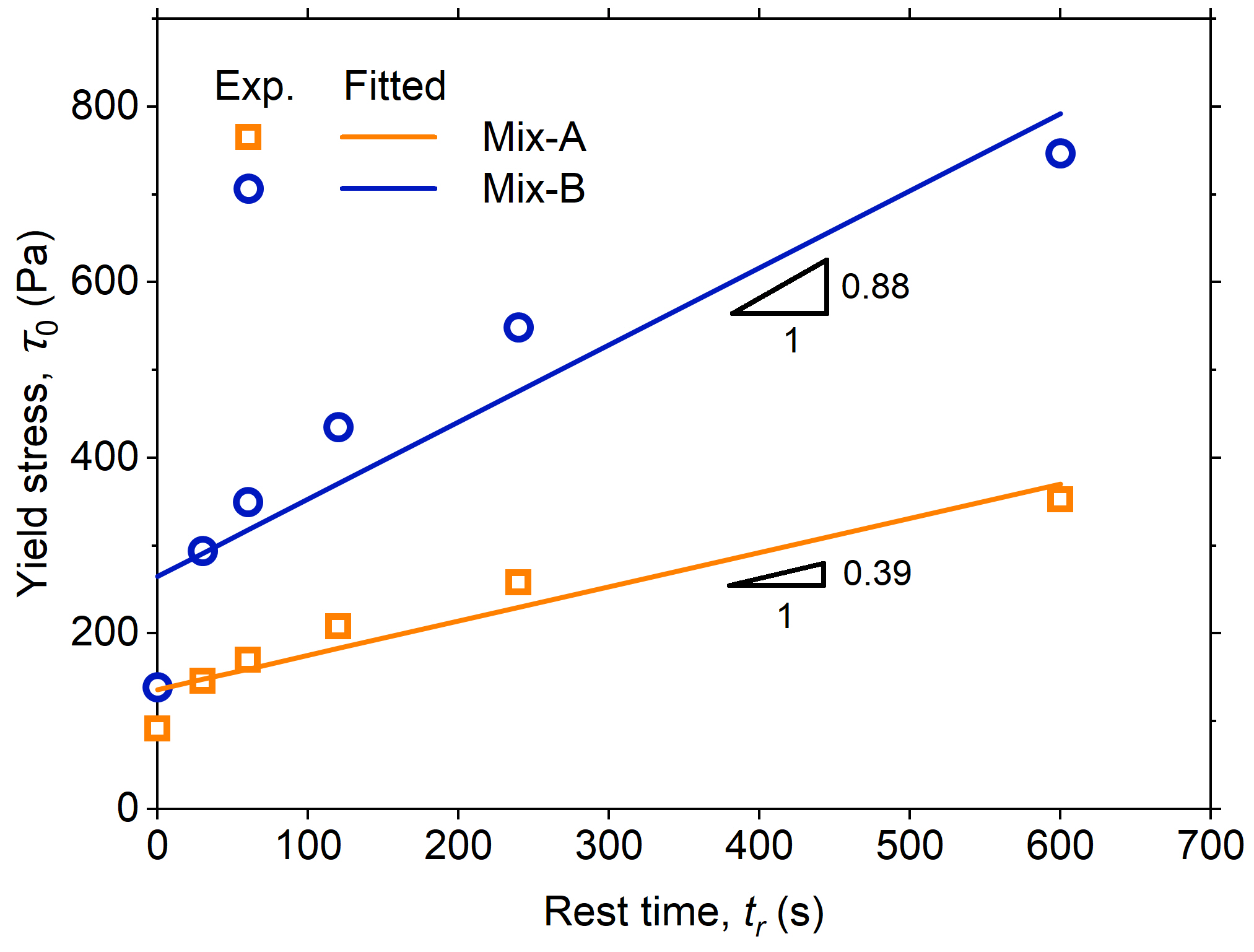}
    \caption{Thixotropic slump flow: Calculation of the flocculation parameter $A_{\mathrm{thix}}$ by linear regression of effective yield stress data. }
    \label{fig: Chris-SF-0}
\end{figure}

To demonstrate the capability of the developed elasto-viscoplastic model in simulating thixotropic concrete, the results of numerical simulations are compared against experimental slump flow tests performed on-site on production mixes. {Yield stress}, plastic viscosity, and slump flow (SF) values are obtained from an EFFC and DFI data set report of concrete tests performed on Tremie Concrete from numerous European and North American sites \citep{krankel2018rheology, feys2018testing}. Two typical concrete mixes from this data set, referred to as Mix-A and Mix-B, are chosen based on the range of concrete properties they represent. The flocculation parameter, $A_{\mathrm{thix}}$, is determined by performing a linear regression from the effective yield stress data $\tilde{\tau}_0$ measured at different rest times, as shown in Fig.~\ref{fig: Chris-SF-0}. It is important to emphasize that this linear fit is used solely to derive $A_{\mathrm{thix}}$. In our simulations, we still use the measured values of $\tau_0$ rather than the fitted values (i.e.~92 and 139 Pa for Mix-A and Mix-B, respectively). The flocculation state $\lambda$ can then be calculated from the measured yield stresses, $\lambda = \tilde{\tau}_0/\tau_0-1$. The final fitted $A_{\mathrm{thix}}$ values are approximately 0.39 for Mix-A and 0.88 for Mix-B. According to Roussel's classification criteria \citep{roussel2006}, Mix-A and Mix-B fall under thixotropic concrete (with $0.1<A_{\mathrm{thix}}<0.5$) and highly thixotropic concrete (with $A_{\mathrm{thix}}>0.5$), respectively. The report documents the slump flow diameter of Mix-A and Mix-B at rest times ($t_r$) of 0s, 240s, and 600s, denoted as SF0, SF240, and SF600. Table \ref{table: 2} shows the other model parameters for the two concrete mixes. 

\hl{It should be noted that the parameters $\alpha$ and $\mu_f$ are not provided in the EFFC/DFI report. Typically, these parameters relate to the properties of concrete and therefore require calibration. Considering that the value of $\alpha$ has minimal impact on the simulation results for concrete immediately after mixing (i.e.,~Tests SF0), we calibrated $\mu_f$ using this experimental dataset and set $\alpha=0$ for this purpose. Then, assuming $\mu_f$ is constant for the same concrete mix across different rest times, we calibrated $\alpha$ using the experimental results for SF240 and SF600. Given the unavoidable errors in both experimental and simulation tests, we deemed a gap between  $\text{SF}{\text{mpm}}$ and $\text{SF}{\text{exp}}$ within ±30 mm to be acceptable.}

\hl{For Mix-A and Mix-B, the calibrated values of $\mu_f$ are 0.30 and 0.35, respectively, both of which fall within a reasonable range for friction coefficients. The corresponding calibrated values of $\alpha$ are 0.15 and 0.4.} This trend suggests a positive correlation between the deflocculation rate $\alpha$ and the flocculation rate $A_{\text{thix}}$. Such behavior is consistent with the established understanding of thixotropy in cementitious materials, in which structural build-up at rest and structural breakdown under shear arise from the same reversible flocculated microstructure. Experimental studies have shown that mixtures exhibiting stronger or faster flocculation tend to form larger but mechanically weaker flocs, which are more susceptible to shear-induced breakage, and that the extent of structural breakdown is closely linked to the subsequent capacity for structural rebuild \citep{roussel2012, ferron2013aggregation, Ye2024showing}.

\begin{table}[h!]
    \caption{Thixotropic slump flow: Summary of test conditions, material parameters, and measured slump flow diameter in experiments and MPM simulations.}
    \centering
    {
    \footnotesize
    \begin{tabular*}{\textwidth}{@{\extracolsep{\fill}}ccccccccccccc@{}}
        \toprule
        Mix ID & Test & $t_r$ & $\rho$ & {$\tilde{\tau}_0$} & {${\tau}_0$} & $A_{\mathrm{thix}}$ & $\lambda$ & $\alpha$ & $\eta$ & $\mu_f$ & $\text{SF}_{\text{exp}}$ & $\text{SF}_{\text{mpm}}$ \\ 
        & & (s) & $(\text{kg/m}^3)$ & (Pa) & (Pa) & (Pa/s) & - & - & (Pa s) & - & (mm) & (mm)\\ 
        \midrule
        \multirow{3}{*}{Mix-A} & SF0 & 0 & \multirow{3}{*}{2400} & 92 & \multirow{3}{*}{92} & \multirow{3}{*}{0.39} & 0 & \multirow{3}{*}{\makecell{0.15*}} & \multirow{3}{*}{19.5} & \multirow{3}{*}{0.30*} & 535 & 532 \\
        & SF240 & 240 &  & 258 &  &  & 1.8 &  &  &  & 500 & 493 \\
        & SF600 & 600 &  & 353 &  &  & 2.8 &  &  &  & 460 & 472 \\
        \midrule
        \multirow{3}{*}{Mix-B} & SF0 & 0 & \multirow{3}{*}{2400} & 139 & \multirow{3}{*}{139} & \multirow{3}{*}{0.88} & 0 & \multirow{3}{*}{\makecell{0.40*}} & \multirow{3}{*}{29} & \multirow{3}{*}{0.35*} & 455 & 459 \\
        & SF240 & 240 &  & 549 &  &  & 2.9 &  &  &  & 430 & 420 \\
        & SF600 & 600 &  & 747 &  &  & 4.4 &  &  &  & 420 & 395 \\        
        \bottomrule
    \end{tabular*}
    }
    \begin{tablenotes}
    \item[*] \footnotesize * Calibrated value. 
    \item[*] \footnotesize Note: $\text{SF}_{\text{exp}}$ - experimental slump flow diameter; and $\text{SF}{\text{mpm}}$ - simulated slump flow diameter by MPM using the elasto-viscoplastic model. 
    \end{tablenotes}
\label{table: 2}
\end{table}

\begin{figure}[h!]
    \centering
    \includegraphics[width=1\linewidth]{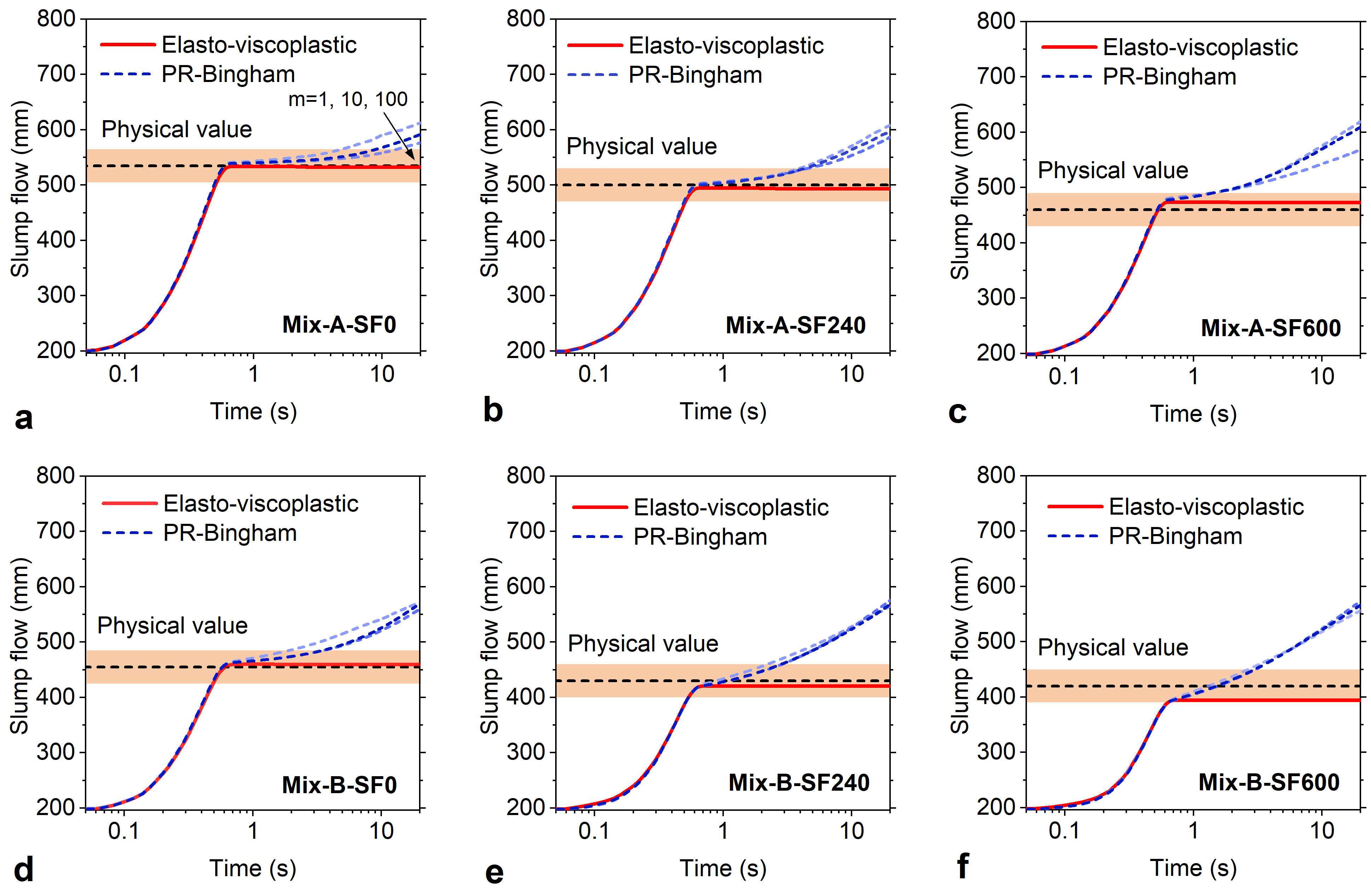}
    \caption{Thixotropic slump flow: Comparison of simulated slump flow and the reported physical value. (a-c) Mix-A and (d-f) Mix-B. The orange-colored rectangular region represents the range of the physical SF±30 mm. \hl{The arrows indicate the results obtained with increasing values of the regularization parameter, i.e., $m=1$, 10, and 100, for the PR-Bingham fluid model. The darker blue dashed curve represents the default case, computed using $m=10$.}}
    \label{fig: Chris-SF-1}
\end{figure}

Fig.~\ref{fig: Chris-SF-1} compares the slump flow obtained from MPM simulations using the elasto-viscoplastic solid model and the PR-Bingham fluid model against experimental measurements. For the solid model, employing a unified set of parameters (${\tau}_0$, $A_{\mathrm{thix}}$, and $\alpha$) generally reproduces the physical test results across various thixotropic states (from SF0 to SF600). Specifically, for Mix-A, the simulated slump flow values closely align with experimental data ($\text{SF}_{\text{exp}}$), with deviations within $\pm 15$ mm. Additionally, the {decrease} in slump flow (i.e.~$\Delta$SF) at rest durations of 240 s and 600 s is also well captured: the experimental results are 35 mm and 75 mm, while the simulated values are 39 mm and 60 mm, respectively.

\begin{figure}[h!]
    \centering
    \includegraphics[width=0.9\linewidth]{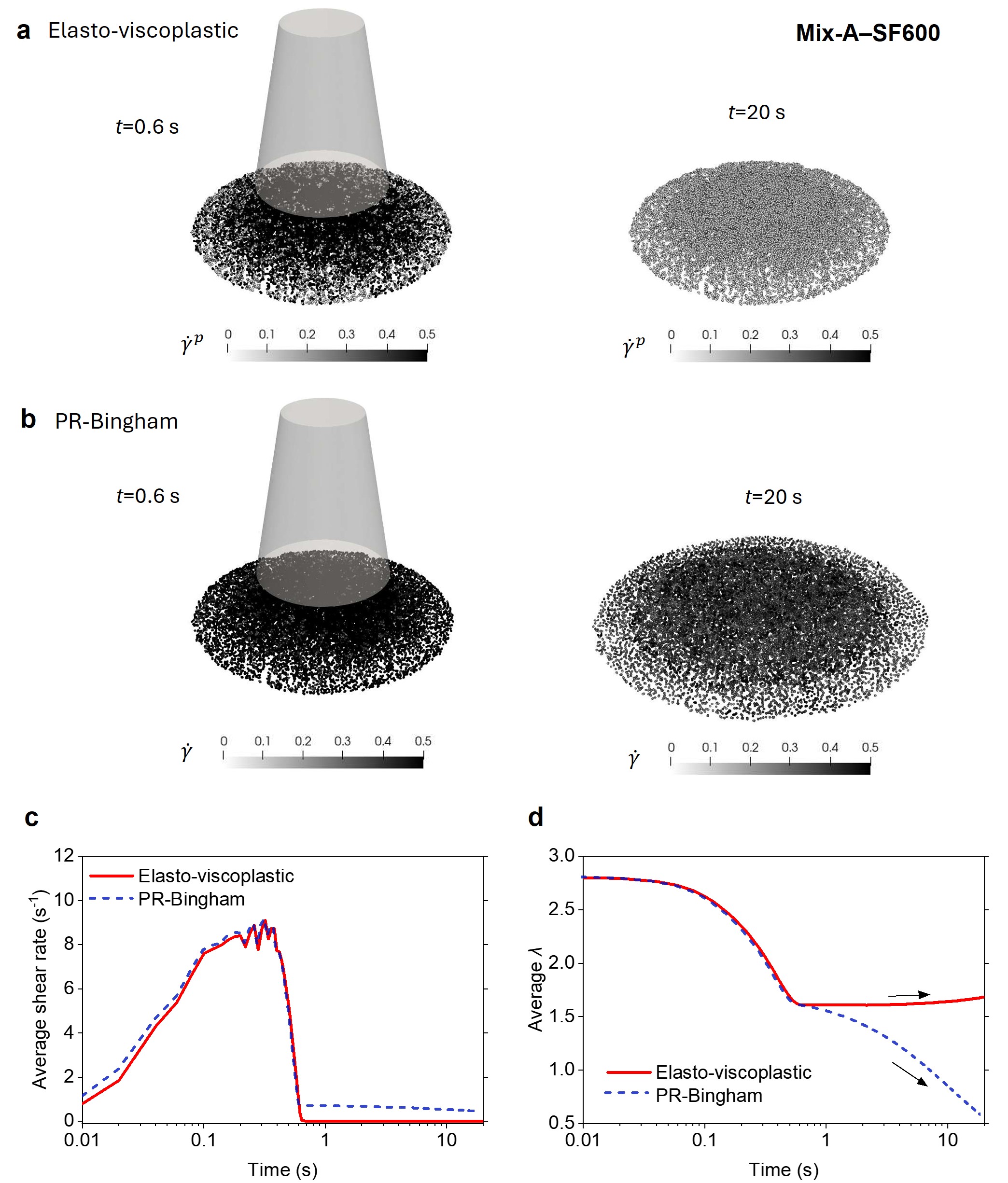}
    \caption{Thixotropic slump flow: Comparison of the simulation results of {Mix-A-SF600} based on the elasto-viscoplastic solid model and the PR-Bingham fluid model. {Contours of (a) plastic shear rate ($\dot\gamma^p$) for the elasto-viscoplastic model and (b) shear rate ($\dot\gamma$) for the PR-Bingham model at $t=$ 0.6 s and 20 s; and evolution of (c) average shear rate ($\dot\gamma^p$ or $\dot\gamma$) and (d) flocculation state variable $\lambda$. The arrows indicate opposite evolution trends of $\lambda$ for the fluid and solid models as time progresses.}}
    \label{fig: Chris-SF-2}
\end{figure}

\begin{figure}[h!]
    \centering
    \includegraphics[width=0.9\linewidth]{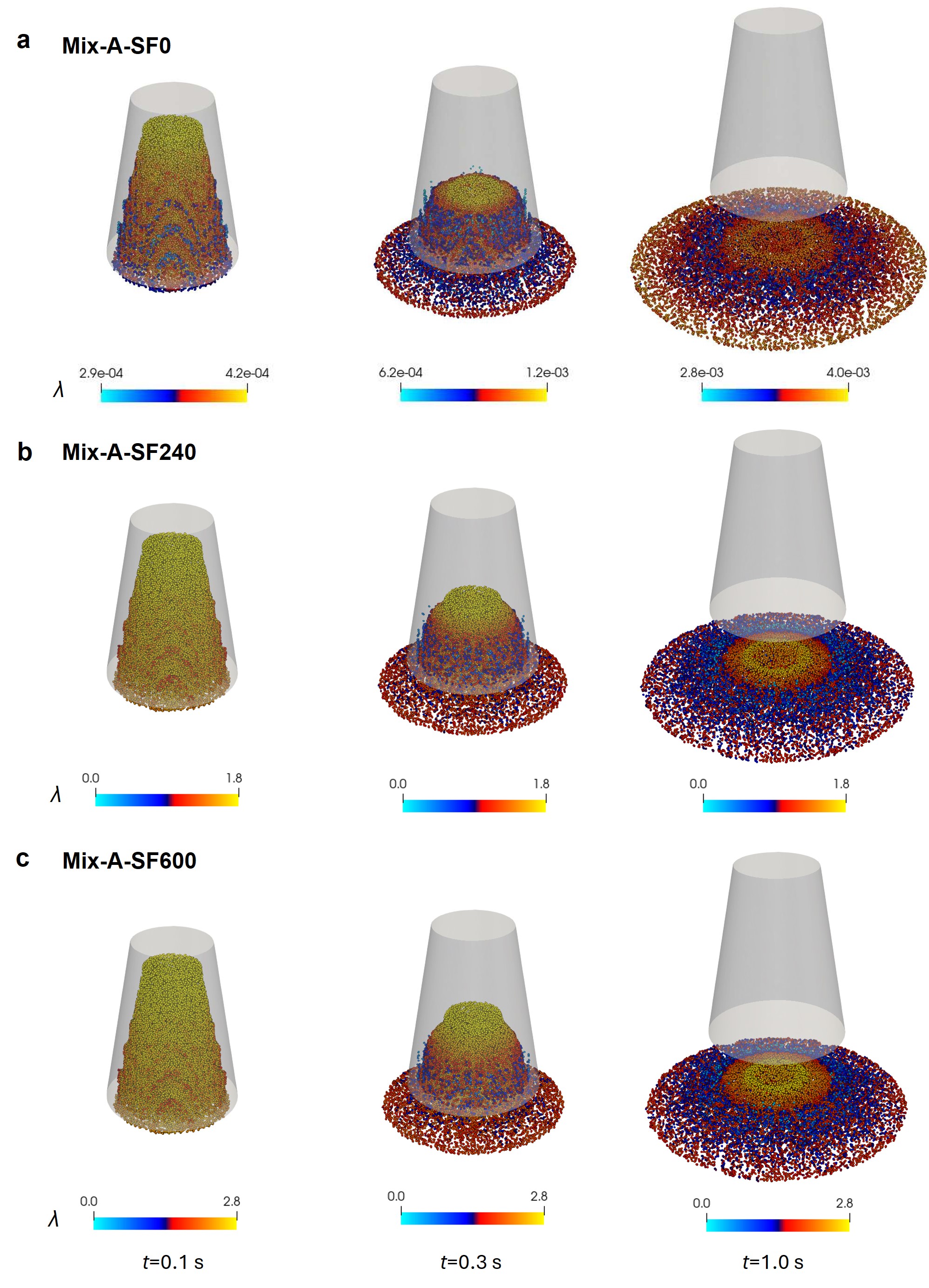}
    \caption{Thixotropic slump flow: Simulation results for Mix-A using the elasto-viscoplastic model at different time instants -- (a) SF0, (b) SF240, and (c) SF600. The contours represent the values of the flocculation state variable $\lambda$.}
    \label{fig: Chris-SF-3}
\end{figure}

\begin{figure}[h!]
    \centering
    \includegraphics[width=0.9\linewidth]{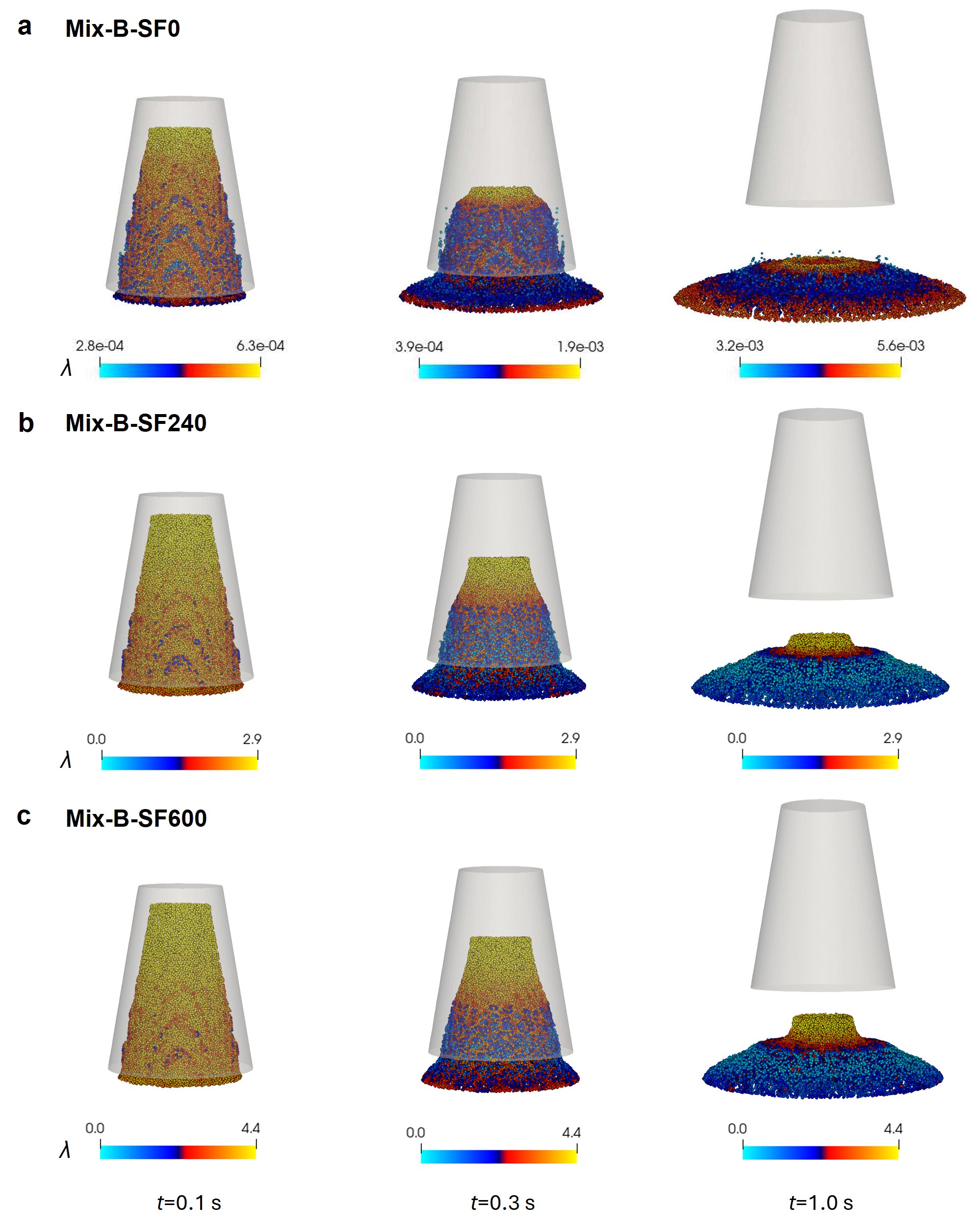}
    \caption{Thixotropic slump flow: Simulation results for Mix-B using the elasto-viscoplastic model at different time instants -- (a) SF0, (b) SF240, and (c) SF600. The contours represent the values of the flocculation state variable $\lambda$.}
    \label{fig: Chris-SF-4}
\end{figure}

Despite its simplicity, the solid model effectively simulates the cessation of concrete flow, a critical behavior that the fluid model fails to capture. As illustrated in Fig.~\ref{fig: Chris-SF-1}, while both models produce nearly identical results during the initial stage, the fluid model predicts continuous flow, resulting in a non-physical overestimation of SF distance at 20 s\hl{, even with a large value of the regularization parameter, e.g., $m=100$}. Figs.~\ref{fig: Chris-SF-2}a and b show contour plots of the flocculation state variable $\lambda$ for Mix-A at 0.6 s and 20 s, for both models. The corresponding temporal evolutions of $\lambda$ and the shear rate $\dot\gamma$ or $\dot\gamma^p$ are presented in Figs.~\ref{fig: Chris-SF-2}c and d. These results indicate that, while both models behave similarly at 0.6 s, the fluid model sustains a low but non-zero shear rate, causing $\lambda$ to decrease continuously and flow to persist. In contrast, the solid model shows the shear rate dropping to zero, accompanied by an increase in $\lambda$, indicating flow cessation and a transition to flocculation. This highlights the clear advantage of the proposed elasto-viscoplastic solid model in simulating concrete flow behavior. For illustrative purposes, Fig.~\ref{fig: Chris-SF-3} and \ref{fig: Chris-SF-4} depict the flow profiles of Mix-A and Mix-B at selected time instants.

\subsection{L-box test}
\label{sec: L-box test}
\begin{figure}[h!]
    \centering
    \includegraphics[width=0.55\linewidth]{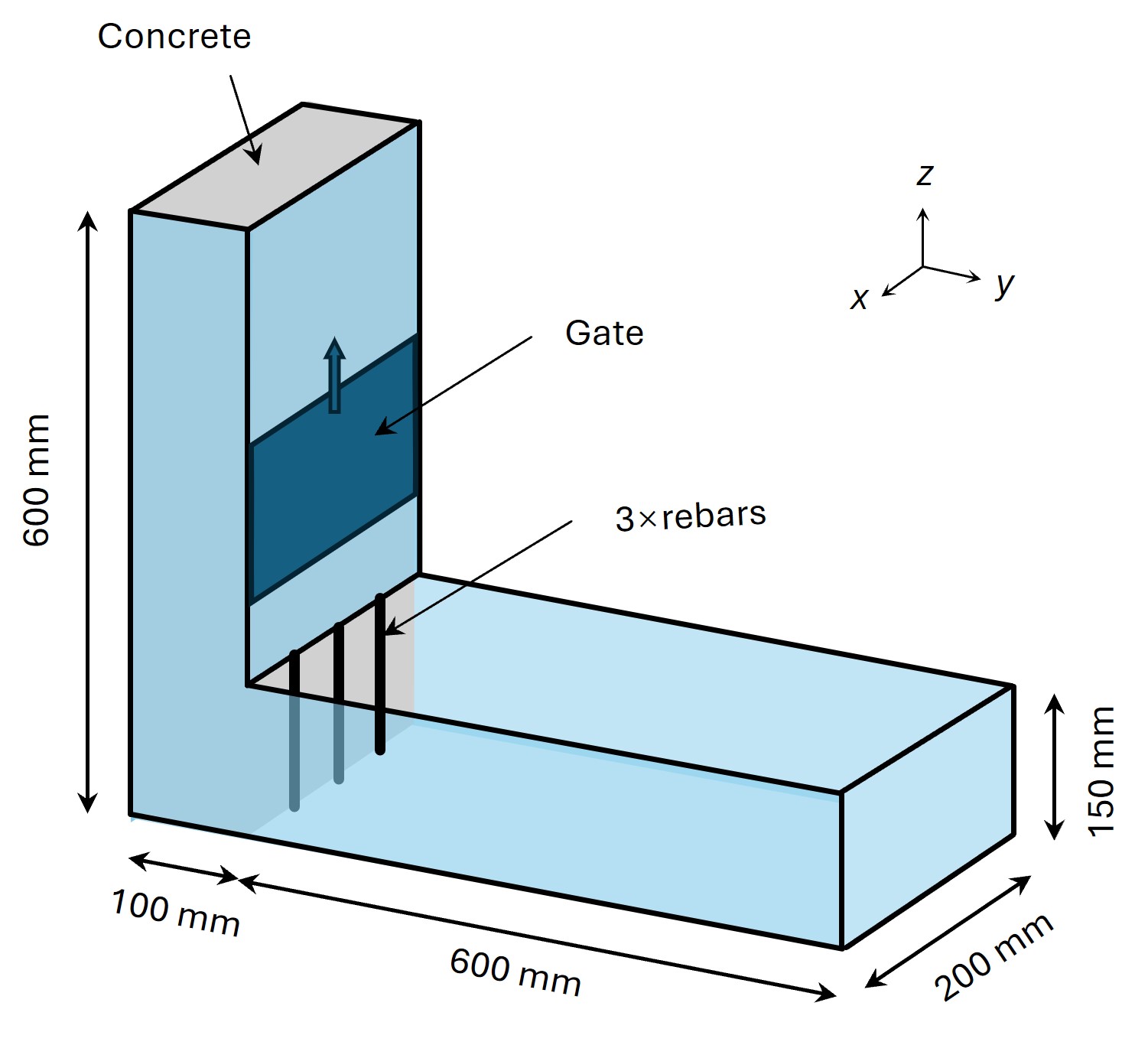}
    \caption{L-box test: Illustration of the geometry and model setup.}
    \label{fig: Chris-L-Box-1}
\end{figure}

The L-Box test \citep{BS1235010} is another widely adopted experiment for evaluating the flowability and passing ability of fresh concrete \citep{nguyen2006correlation}. This test observes the flow behavior of concrete in confined spaces, particularly through reinforcing bars, to assess its workability. \hl{Despite the challenge for continuum-based models to resolve coarse aggregate blocking, the L-Box test offers a well-defined benchmark with spatially varying shear rates and a sudden constriction. This allows us to evaluate the model's ability to reproduce bulk flow behavior, filling height/flow distance, and local flocculation-state evolution as the material is sheared through the confined rebar gaps.} As shown in Fig.~\ref{fig: Chris-L-Box-1}, the L-Box apparatus consists of a concrete reservoir (200 mm × 100 mm × 600 mm), a rebar grid (3 {$\times$ 12 mm $\Phi$ rebars}), and an end box (200 mm × 150 mm × 600 mm). At the beginning of the test, 12 liters of concrete are poured into the reservoir and allowed to rest in an initial stationary state. Subsequently, the gate is opened, allowing the concrete to flow under gravity through the rebar grid and toward the end box. 

Depending on the flow behavior of the concrete, the test results can be categorized into two typical cases:
\begin{enumerate}
    \item Complete flow -- concrete flow reaches the end of the end box. In this case, the time taken for the concrete to reach the end (denoted as $t^{\mathrm{end}}$) is recorded, along with the final height of the concrete at the end of the box after the flow has stopped (denoted as $h$). The two parameters reflect the flow velocity and the passing ability of the concrete through the rebar grid.
    \item Incomplete flow -- concrete flow does not reach the end of the box due to excessive viscosity or insufficient flowability. In this scenario, the final flow distance of the concrete (denoted as $d$) is measured, and the time at which the concrete ceases flow (denoted as $t^{\mathrm{final}}$) is also recorded to evaluate the extent of the flow and the location of the stagnation point.
\end{enumerate}

To simulate the L-box test, the entire computational domain is divided into uniform hexahedral grids. In the x-direction, three rebars are evenly placed, and the grid size is largely determined by the dimensions and positions of the rebars. Because accurately simulating the circular cross-section of the rebars is challenging, their cross-sections are simplified to rectangular shapes with a width equal to one grid cell. Considering that the rebars divide the L-box into four equal parts along the x-direction, 19 cells are created along this direction, each with a width of 10.5 mm. In the y- and z-directions, a uniform grid size of 10 mm is used. As a result, the dimensions of each grid cell are 10.5 × 10 × 10 mm. Due to the irregularity of the side lengths, isoparametric elements are considered. Each grid cell contains 8 material points, resulting in a total of {91,200} material points, ensuring high resolution and computational accuracy. The quasi-static stress under the self-weight of the concrete is adopted as the initial stress condition. Additionally, for computational simplicity, we assume the gate-opening time is negligible; thus, we do not explicitly simulate the gate-opening process. \hl{We further simplify the interface condition between fresh concrete and rebars by merely assuming the rebars to be completely rough in all directions. Consequently, the rebar constraints can be represented as Dirichlet boundary conditions, achieved by setting the nodal kinematics to zero in MPM.} Similar to the previous section, we simulate the flow behavior of two concrete mixes, Mix-A and Mix-B, at two different rest times ($t_r=$ 0 s and 600 s). 

\begin{table}[h!]
    \caption{L-box test: Summary of test conditions, material parameters, and experimental/simulation results.}
    \centering
    {
    \footnotesize
    \begin{tabular*}{\textwidth}{@{\extracolsep{\fill}}ccccccccccccccc@{}}
        \toprule
        Mix ID & Test & $t_r$ & {${\tau}_0$} & $A_{\mathrm{thix}}$ & $\lambda$ & $\alpha$ & $\eta$ & $\mu_f$ & $h_{\text{exp}}$ & $h_{\text{mpm}}$ & $t^{\mathrm{end}}_{\text{exp}}$ & $t^{\mathrm{end}}_{\text{mpm}}$ \\ 
        & & (s) & (Pa) & (Pa/s) & - & -  & (Pa s) & - & (mm) & (mm) & (s) & (s)\\ 
        \midrule
        \multirow{2}{*}{Mix-A} & SF0 & 0 & \multirow{2}{*}{92} & \multirow{2}{*}{0.39} & 0 & \multirow{2}{*}{{$0.15^\dagger$}} & \multirow{2}{*}{19.5} & \multirow{2}{*}{0.25*} & 70 & 50 & 1.4 & 1.1  \\
        & SF600 & 600 &  &  & 2.8 &  &  &  & 45 &  31 & 2.6 & 2.5 \\      
        \bottomrule
        \toprule
        Mix ID & Test & $t_r$ & {${\tau}_0$} & $A_{\mathrm{thix}}$ & $\lambda$ & $\alpha$ & $\eta$ & $\mu_f$ & $d_{\text{exp}}$ & $d_{\text{mpm}}$ & $t^{\mathrm{final}}_{\text{exp}}$ & $t^{\mathrm{final}}_{\text{mpm}}$ \\ 
        & & (s) & (Pa) & (Pa/s) & - & - & (Pa s) & - & (mm) & (mm) & (s) & (s)\\ 
        \midrule
        \multirow{2}{*}{Mix-B} & SF0 & 0 & \multirow{2}{*}{139} & \multirow{2}{*}{0.88} & 0 & \multirow{2}{*}{{$0.40^\dagger$}} & \multirow{2}{*}{29} & \multirow{2}{*}{0.45*} & 580 & 590 & 11.5 & 9 \\
        & SF600 & 600 &  &  & 4.4 &  &  &  & 520 & 520 & 13 & 10 \\        
        \bottomrule
    \end{tabular*}
    }
    \begin{tablenotes}
    \item[*] \footnotesize * Calibrated value. 
    \item[*] \footnotesize {$\dagger$ $\alpha$ is taken as the value calibrated from the slump flow test in Section \ref{sec: slump flow test}.}
    \item[*] \footnotesize Note: $h$ and  $t^{\mathrm{end}}$ are the height of the concrete at the end of the box and the time at which the concrete reaches the end, respectively; $d$ and $t^{\mathrm{final}}$ denote the final flow distance and the time the concrete ceases to flow, respectively; and $\Box_{\rm exp}$ and $\Box_{\rm mpm}$ denote the experimental and simulated results using the elasto-viscoplastic model, respectively.
    \end{tablenotes}
\label{table: 3}
\end{table}

The experimental data, recorded from the EFFC and DFI datasets \citep{krankel2018rheology, feys2018testing}, served as a reference for validating the numerical model. The experimental results indicate that Mix-A is a complete flow, whereas Mix-B is an incomplete flow, representing two distinct flow behaviors. Table \ref{table: 3} presents the material parameters and key experimental results for reference. \hl{It should be noted that, since the same concrete mix as in the slump flow test is used, the previously calibrated deflocculation parameter $\alpha$ is adopted for consistency (they are not recalibrated for this geometry). However, the friction coefficient $\mu_f$, being a property related to the interaction between the concrete and the contact interface, should be recalibrated. As with the slump flow test, we use tests with $ t_r=0$ to calibrate $\mu_f$ for the two mixes. Here, we assume that the coefficient of friction is identical for all boundary walls, including the sides and the bottom. Finally, the calibrated $\mu_f$ is 0.25 and 0.45 for Mix-A and Mix-B, respectively (see Table \ref{table: 3}).} Table \ref{table: 3} also presents key metrics measured in both the experiments and simulations, including $t^{\mathrm{end}}$ and $h$ for Mix-A, as well as $t^{\mathrm{final}}$ and $h$ for Mix-B. It can be observed that the $\alpha$ values calibrated using the slump flow tests can also be successfully applied to the L-box simulations, yielding results that align well with the experimental data. 

\begin{figure}[h!]
    \centering
    \includegraphics[width=1\linewidth]{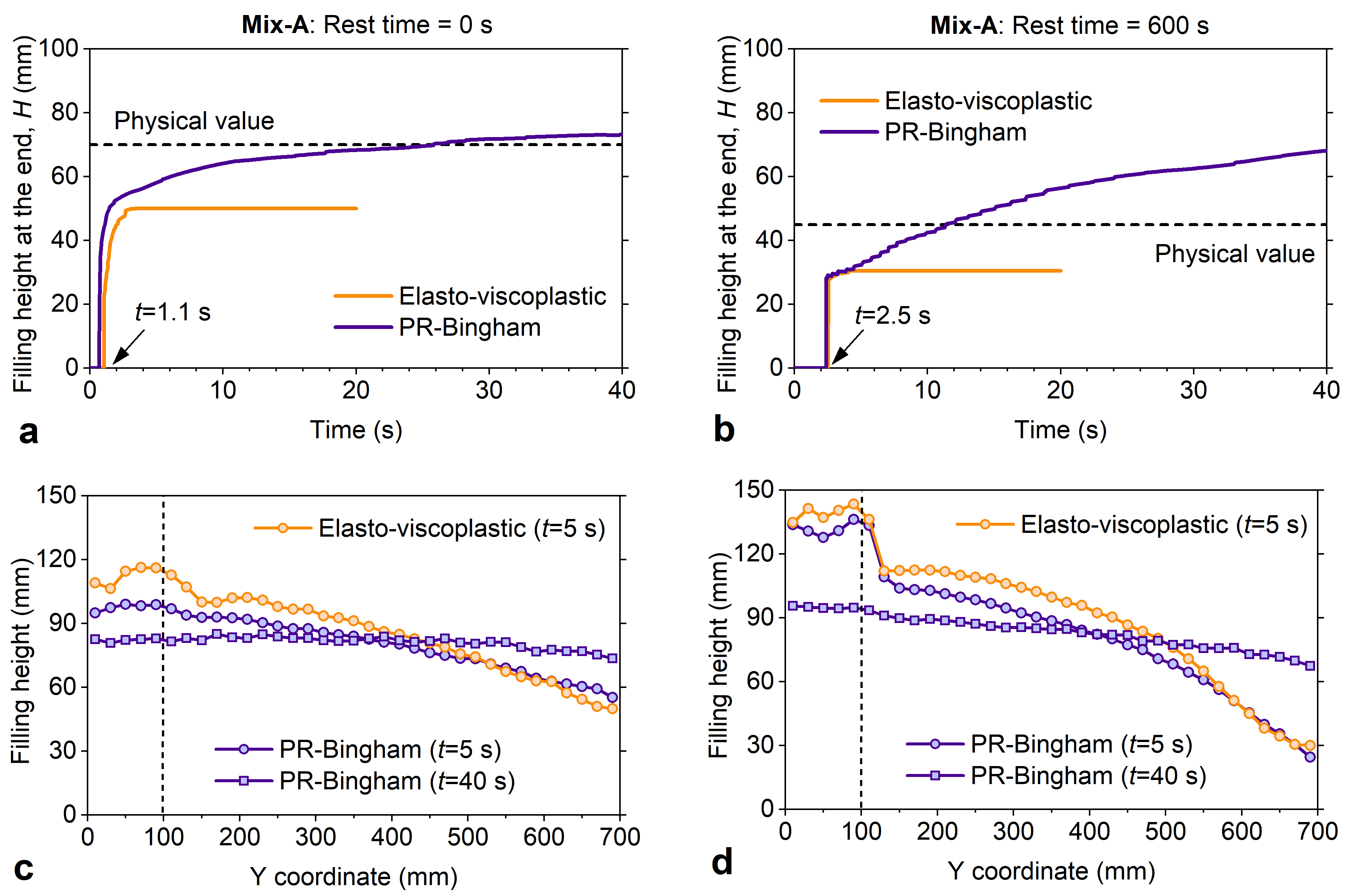}
    \caption{L-box test: Simulation results of Mix-A: (a-b) filling height at the end of the box; (c-d) surface configuration (along the central y-z plane) at different time instants. The left and right columns represent the results for rest times $t_r = 0$ and \hl{$t_r = 600$ s}, respectively. The vertical dashed lines show where the L-box transitions from the concrete reservoir to the end box.}
    \label{fig: Chris-L-Box-2}
\end{figure}

In this study, both Mix-A and Mix-B are simulated using the PR-Bingham fluid model using the same parameters, and the results are compared with those obtained from the elasto-viscoplastic solid model. Figs.~\ref{fig: Chris-L-Box-2}a and b compare the evolution of the filling height of concrete at the end of the L-box for Mix-A with rest times of 0 s and \hl{600 s}, respectively. In both tests, the elasto-viscoplastic solid model and the PR-Bingham fluid model reached the end of the box at approximately the same time. However, the solid model ceased flowing after just 2 seconds, while the fluid model continued flowing for up to 40 seconds. Although the fluid model results in Fig.~\ref{fig: Chris-L-Box-2}a appear to align more closely with the experimental data after the initial stage, this trend is unphysical, as the values keep increasing over time due to creeping motion. In Fig.~\ref{fig: Chris-L-Box-2}b, the fluid model’s results at 40 seconds significantly exceed the experimental values, and it is evident that this discrepancy would continue to increase if the simulation were extended further. Figs.~\ref{fig: Chris-L-Box-2}c and d compare the free surface configurations of Mix-A at the end of the simulations. For the solid model, despite being simulated for only 20 seconds, the flow had completely stopped, resulting in a distinct sloped free surface. In contrast, the fluid model, even after 40 seconds, produced a nearly flat free surface.

\begin{figure}[h!]
    \centering
    \includegraphics[width=1\linewidth]{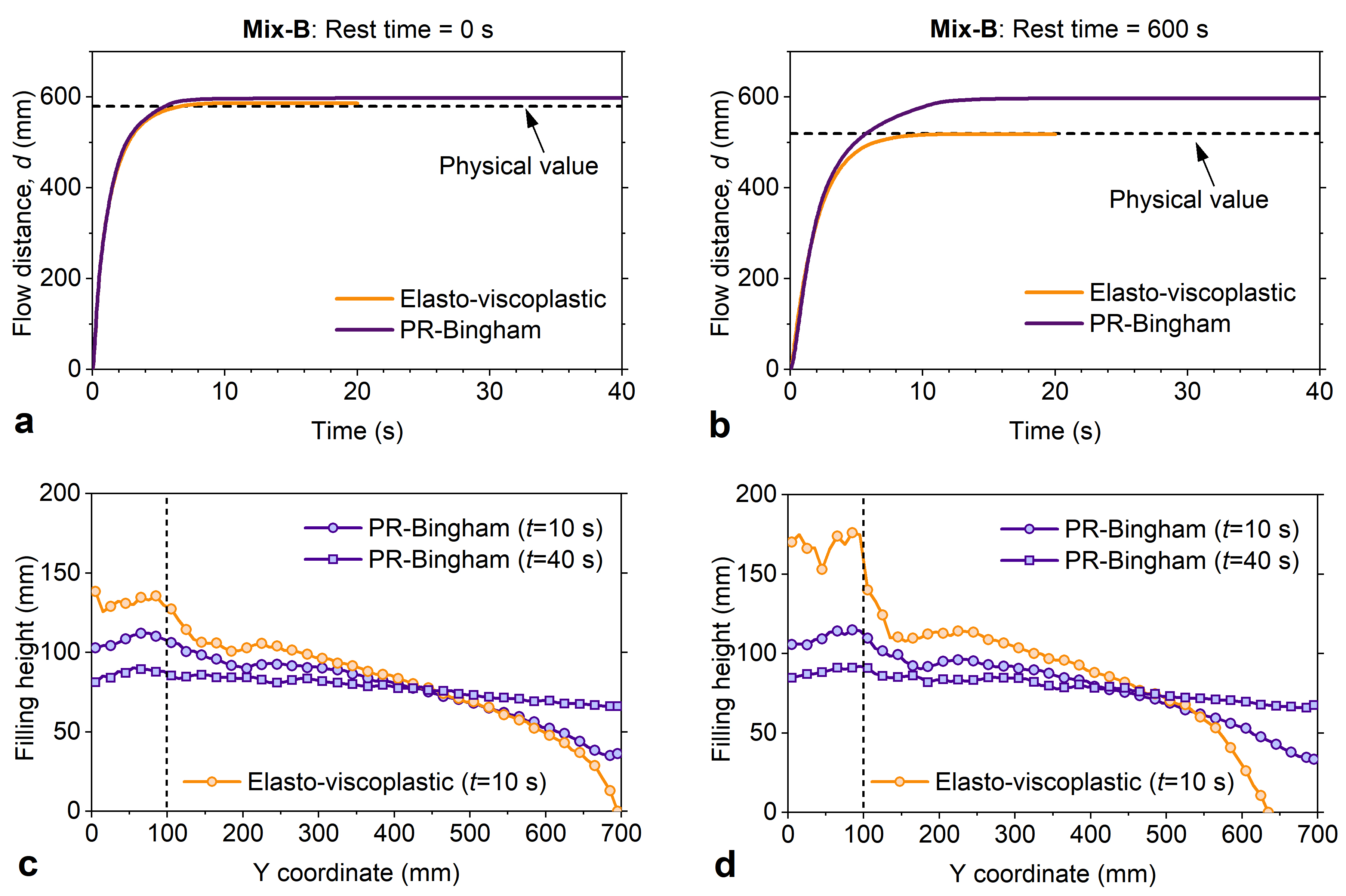}
    \caption{L-box test: Simulation results of Mix-B: (a-b) flow distance $d$ measured from the transition between the concrete reservoir and the end box; (c-d) surface configuration (along the central y-z plane) at different time instants. The left and right columns represent the results for rest times $t_r = 0$ and $t_r = 600$ s, respectively. The vertical dashed lines show where the L-box transitions from the concrete reservoir to the end box.}
    \label{fig: Chris-L-Box-3}
\end{figure}

A similar pattern is observed for Mix-B. Since Mix-B did not reach the end of the box before stopping flowing, Figs.~\ref{fig: Chris-L-Box-3}a and b compare the flow distance instead. In the early stages of flow, the solid model and fluid model produced nearly identical flow fronts. However, the solid model stopped flowing after reaching distances of 590 mm and 520 mm for the respective tests, which matched the experimental results very well. Conversely, the fluid model continued flowing and eventually reached the end of the box, which is inconsistent with the experimental observations. Examining the final concrete shapes in Figs.~\ref{fig: Chris-L-Box-3}c and d, the fluid model produced a nearly horizontal surface at 40 seconds. The inability to maintain the concrete's shape introduces significant distortions in the simulation results, with the degree of inaccuracy increasing as the simulation progresses. 

\hl{It is worth noting that the simulated filling heights for Mix-A (Figs.~\ref{fig: Chris-L-Box-2}a and b; Table~3) are apparently lower than the experimental measurements, indicating that the model overpredicts flow resistance. Our primary hypothesis is the simplified boundary condition: a single friction coefficient $\mu_f$ is applied to all walls, even though the vertical sidewalls likely offer much lower resistance than the bottom plate in reality. This uniform approximation thus overestimates the retarding effect on these surfaces and decelerates the simulated flow front. In this specific case, $\mu_f$ is calibrated by balancing the concrete height at the box end, $h$, with the time it takes for the concrete to reach that end, $t^{\mathrm{end}}$. When $\mu_f$ is small, the flow accelerates, leading to a shorter $t^{\mathrm{end}}$ and a larger recorded height $h$. Conversely, increasing $\mu_f$ increases the flow resistance, which prolongs $t^{\mathrm{end}}$ and consequently reduces the final height $h$. Because the experimentally measured $t^{\mathrm{end}}$ is relatively short -- only a few seconds for Mix-A -- we decided to optimize $\mu_f$ mainly to reproduce the observed time. In addition, the incomplete knowledge of how to model the interaction between fresh concrete and rebars at their interface may be a significant contributing factor. Currently, we assume a fully rough interface that may restrict the motion of the bulk concrete material. Moreover, the exclusion of coarse aggregates in the continuum model may also have a smaller, yet non-negligible, impact, as their presence can generate extra localized shear-thinning effects that facilitate flow. Nevertheless, the model successfully reproduces the essential qualitative trends, including flow cessation and the relative difference between the two rest times.}

\begin{figure}[h!]
    \centering
    \includegraphics[width=0.88\linewidth]{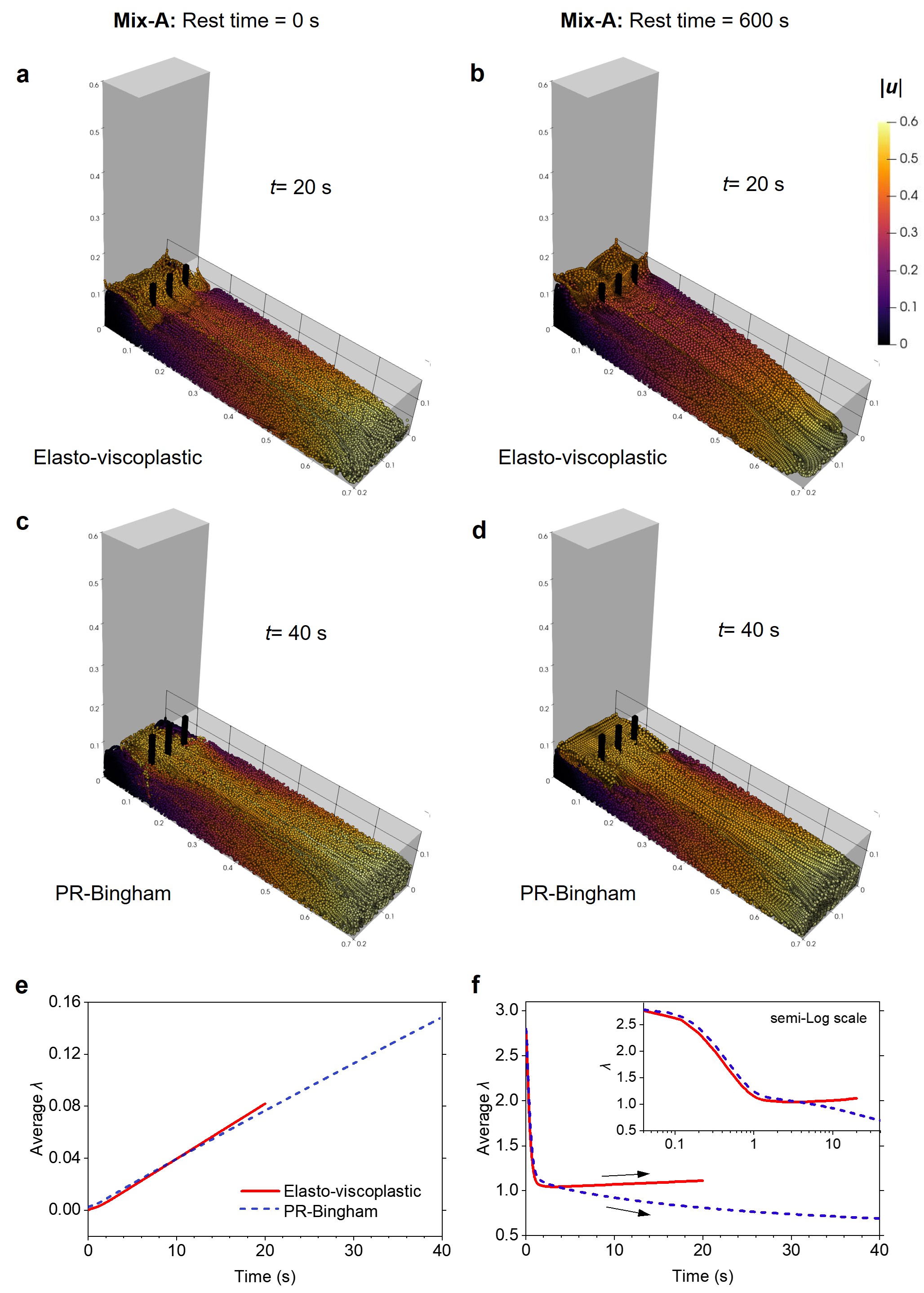}
    \caption{L-box test: Simulation results of Mix-A. (a-b) final concrete configuration simulated by the elasto-viscoplastic solid model; (c-d) final concrete configuration simulated by the PR-Bingham fluid model; (e-f) evolution of the flocculation state $\lambda$ (calculated as the average value across all material points). The left and right columns represent the results for rest times $t_r = 0$ and \hl{$t_r = 600$ s}, respectively. {The arrows show opposite evolution trends of $\lambda$ of the fluid and solid models as time progresses.}}
    \label{fig: Chris-L-Box-4}
\end{figure}

\begin{figure}[h!]
    \centering
    \includegraphics[width=0.88\linewidth]{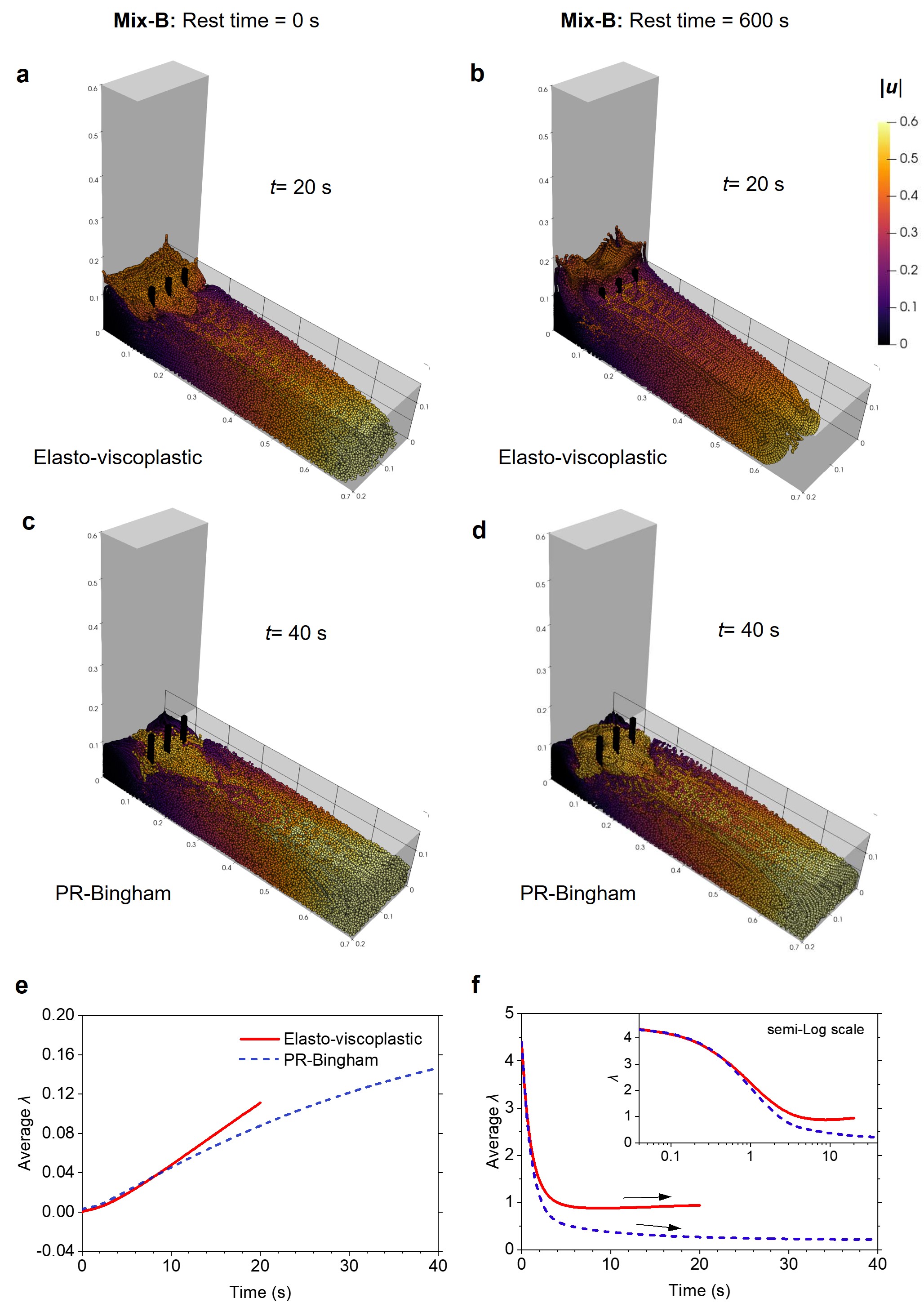}
    \caption{L-box test: Simulation results of Mix-B: (a-b) final concrete configuration simulated by the elasto-viscoplastic solid model; (c-d) final concrete configuration simulated by the PR-Bingham fluid model; (e-f) evolution of the flocculation state $\lambda$ (calculated as the average value across all material points). The left and right columns represent the results for rest times $t_r = 0$ and \hl{$t_r = 600$ s}, respectively. {The arrows show opposite evolution trends of $\lambda$ of the fluid and solid models as time progresses.}}
    \label{fig: Chris-L-Box-5}
\end{figure}

Figs.~\ref{fig: Chris-L-Box-4}a-d and \ref{fig: Chris-L-Box-5}a-d illustrate the final concrete configurations for Mix-A and Mix-B, respectively, as simulated using the solid model and the fluid model. Meanwhile, Figs.~\ref{fig: Chris-L-Box-4}e-f and \ref{fig: Chris-L-Box-5}e-f further compare the evolution of the average flocculation state parameter, $\lambda$, for both models. {For cases with $t_r = 0$, both models exhibit a gradual increase in $\lambda$ over time as flocculation process dominates. However, the $\lambda$ of the fluid model increases at a slower rate compared to the solid model as time progresses, as the deflocculation rate term does not vanish as the fluid model keeps shearing over time. In contrast, for the solid model, the deflocculation rate term diminishes entirely, allowing for a more consistent accumulation of $\lambda$.} For the case with \hl{$t_r = 600$ s}, the initial values of $\lambda$ for both the solid and fluid models are nearly identical. However, the fluid model continues to deflocculate throughout the simulation for the same reason. Meanwhile, the solid model initially deflocculates and exhibits further flocculation after the flow stops. 

\begin{figure}[h!]
    \centering
    \includegraphics[width=1\linewidth]{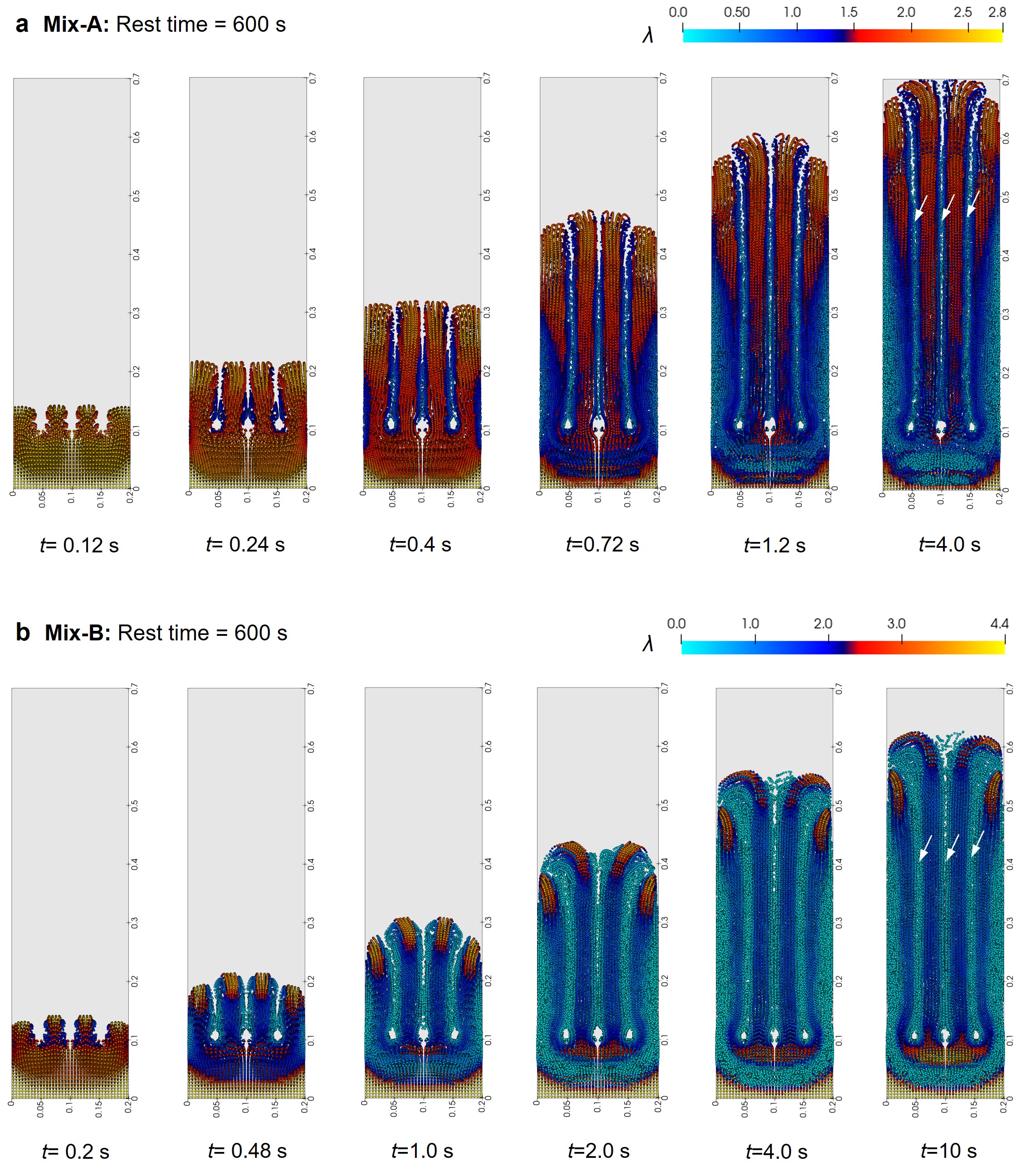}
    \caption{L-box test: Simulation results of the $\lambda$ distribution at the cross-section $z = 0.05$ m: (a) Mix-A at $t=$0.12, 0.24, 0.4, 0.72, 1.2, and 4.0 s, and (b) Mix-B at $t=$ 0.2, 0.48, 1.0, 2.0, 4.0, and 10 s. {The white arrows highlight the three distinct low-$\lambda$ streaks due to the presence of rebars.}}
    \label{fig: Chris-L-Box-6}
\end{figure}

Finally, it is important to discuss the role of rebars. The L-box test is designed to simulate the ability of concrete to flow through narrow passages, such as the gaps between rebars. As concrete flows through the rebar grid, it experiences additional shearing, leading to rapid deflocculation. To observe this phenomenon, we extracted the $\lambda$ profile at $z = 0.05$-m cross section at different time steps. Figs.~\ref{fig: Chris-L-Box-6}a and b show the results for Mix-A and Mix-B, respectively. These figures illustrate the process of concrete being squeezed through the gaps between the rebars and gradually merging downstream. It can be observed that the $\lambda$ values of the concrete flowing near the sides of the rebars are significantly lower than those farther away from the rebars. As the concrete flows forward, three distinct low-$\lambda$ streaks emerge. This finding highlights the significant role of rebars in altering the flow pattern and flocculation state of the concrete. It also underscores the excellent performance of the proposed model in capturing the complex interactions between concrete and boundaries.

\hl{As aforementioned, the present study treats fresh concrete as a homogenized, single-phase continuum, so the current MPM framework does not resolve individual coarse aggregate particles, nor their collective distribution as an aggregate concentration field, and therefore cannot capture aggregate-scale blocking or arching phenomena. This can lead to discrepancies between the L-Box simulation and experimental observations. To address this limitation, a combined MPM–DEM framework—where coarse aggregate particles are modeled explicitly with DEM and the surrounding cement paste or mortar matrix is treated as a continuum using MPM—offers a promising alternative for future studies \citep{jiang2022hybrid, fang2026novel}. Such an approach is expected to provide higher-fidelity predictions of aggregate-scale passing-ability behavior in confined-flow configurations, including the L-Box \citep{BS1235010} and J-Ring \citep{BSEN12350_12_2010} tests.}

\subsection{V-funnel test}
The V-funnel test \citep{BSEN12350_9_2010} is a standard method used to evaluate the viscosity and filling ability of fresh concrete \citep{mu2022simulation, thanh2020numerical, alyhya2017simulation}. It measures the time required for a specific volume of concrete to flow through a narrow, funnel-shaped outlet under the influence of gravity. This test provides critical insights into the rheological properties of the concrete, such as its ability to flow smoothly without external compaction and resistance to segregation. 

\begin{figure}[h!]
    \centering
    \includegraphics[width=0.5\linewidth]{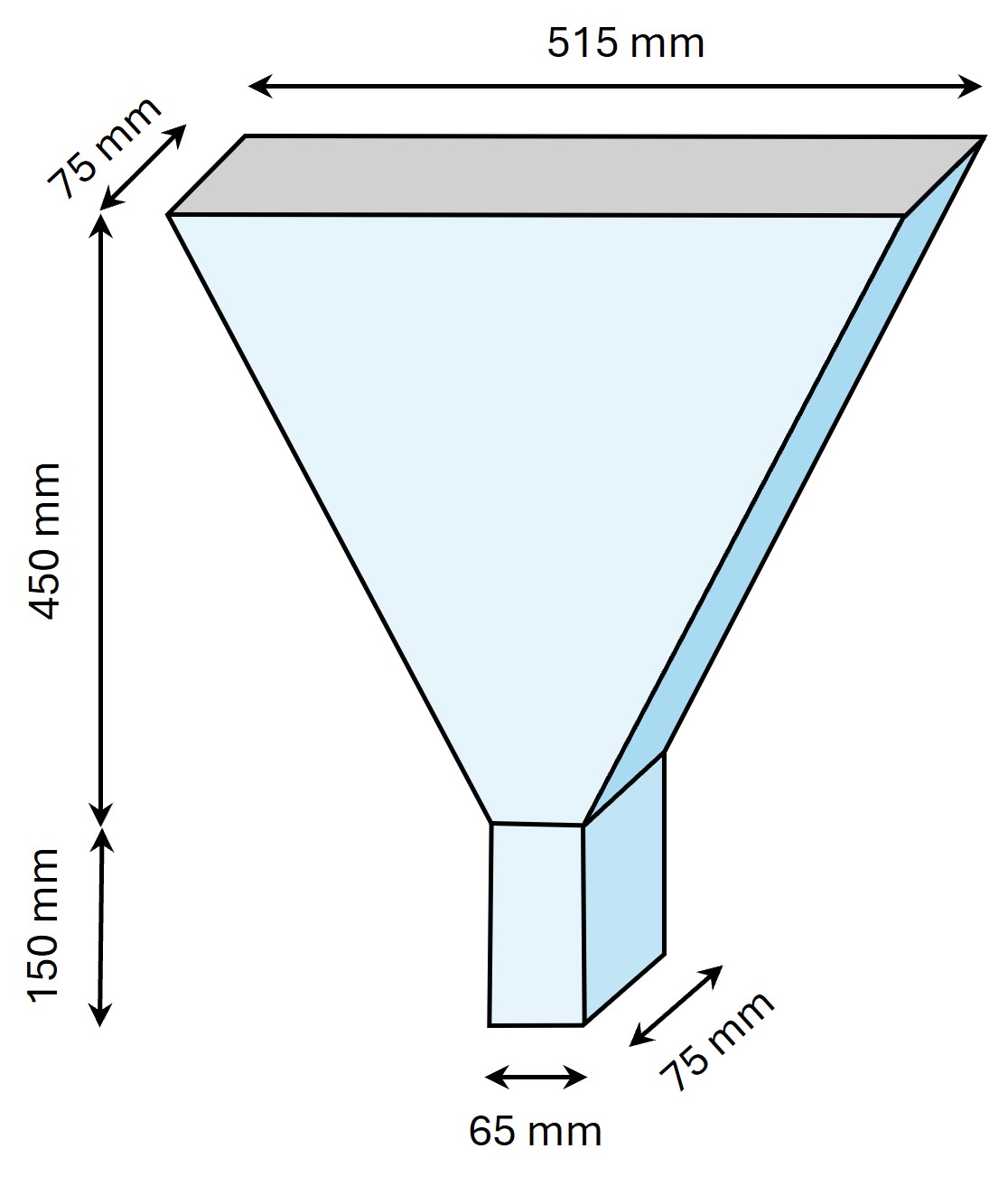}
    \caption{V-funnel test: Illustration of the geometry and model setup.}
    \label{fig: V-funnel-1}
\end{figure}

The standard geometry of the V-funnel, as illustrated in Fig.~\ref{fig: V-funnel-1}, consists of an inclined funnel-shaped container with a height of approximately 600 mm, a top opening of 515 mm × 75 mm, and a rectangular outlet at the bottom measuring 65 mm × 75 mm. The total volume of the funnel is about 12 liters. In the standard lab test procedure, the funnel is first filled with concrete, which is then left to rest for 10 $\pm$ 2 seconds to allow air bubbles to escape and stabilize. The bottom gate is then opened, and the time from gate opening until the interior of the funnel becomes fully visible through the outlet is recorded. This time is defined as the V-funnel flow time ($t_v$). If the concrete flow is discontinuous or blockage occurs, it indicates that the concrete's viscosity is too high, rendering it unsuitable for this test. $t_v$ indirectly characterizes the viscosity and flowability of the concrete. A shorter $t_v$ indicates better flowability, while an excessively long or discontinuous flow suggests high viscosity or insufficient flowability.

\begin{figure}[h!]
    \centering
    \includegraphics[width=1\linewidth]{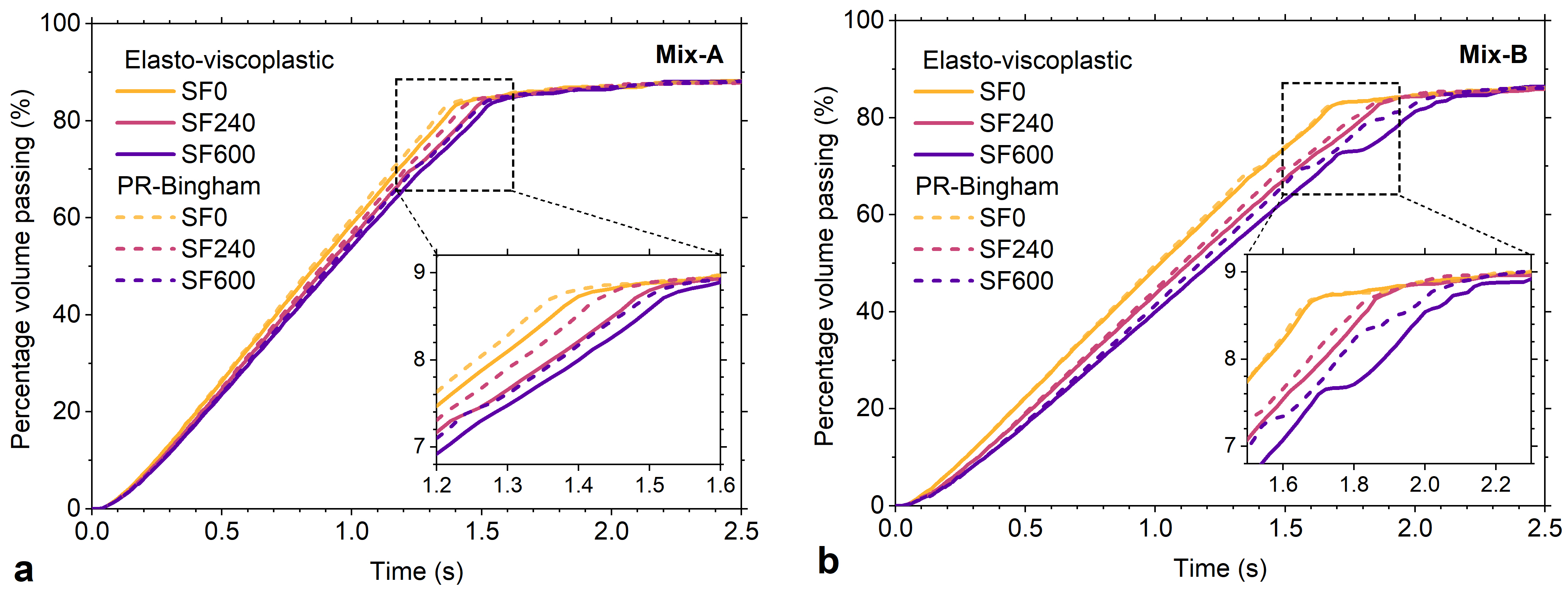}
    \caption{V-funnel test: Evolution of the percentage of concrete volume passing through the bottom gate of the V-funnel for (a) Mix-A and (b) Mix-B at various rest times.}
    \label{fig: V-funnel-2}
\end{figure}

In the MPM simulation, a regular hexahedral grid is utilized as the background mesh, with a grid size of 10.72 mm. Frictional boundary conditions are applied to all walls, while step boundaries are used to discretize the inclined surfaces. The time step size is set to $5 \times 10^{-5}$ s. Simulations are performed for both Mix-A and Mix-B; however, due to the absence of relevant experimental data, validation is not conducted. We continue to use the previously calibrated parameters, except for the friction coefficient, which is fixed at 0.3 for both Mix-A and Mix-B. Both the solid and fluid models are evaluated in the simulations. The measured $t_v$ values for both solid and fluid models are summarized in Table \ref{table: 4}. Additionally, Fig.~\ref{fig: V-funnel-2} illustrates the evolution of the percentage of concrete volume passing through the bottom gate of the V-funnel for Mix-A and Mix-B at various rest times.



\begin{table}[h!]
    \caption{V-funnel test: Summary of test conditions, material parameters, and simulation results.}
    \centering
    {
    \footnotesize
    \begin{tabular*}{\textwidth}{@{\extracolsep{\fill}}ccccccccccc@{}}
        \toprule
       Mix ID & Test & $t_r$ & {${\tau}_0$} & $A_{\rm thix}$ & $\lambda$ & $\alpha$ & $\eta$ & $\mu_f$ & $t_{\rm{v,\, solid}}$ & $t_{\rm{v,\, fluid}}$ \\ 
        & & (s) & (Pa) & (Pa/s) & - & -  & (Pa s) & - & (s) & (s)\\ 
        \midrule
        \multirow{3}{*}{Mix-A} & SF0 & 0 & \multirow{3}{*}{92} & \multirow{3}{*}{0.39} & 0 & \multirow{3}{*}{0.15$^\dagger$} & 
        \multirow{3}{*}{19.5} & \multirow{3}{*}{0.3} & 1.40 & 1.36  \\
        & SF240 & 240 &  &  & 1.8 &  &  &  & 1.50 & 1.46 \\    
        & SF600 & 600 &  &  & 2.8 &  &  &  & 1.54 & 1.50 \\      
        \midrule
        \multirow{3}{*}{Mix-B} & SF0 & 0 & \multirow{3}{*}{139} & \multirow{3}{*}{0.88} & 0 & \multirow{3}{*}{0.40$^\dagger$} & \multirow{3}{*}{29} & \multirow{3}{*}{0.3} & 1.68 & 1.66 \\
        & SF240 & 240 &  &  & 2.9 &  &  &  & 1.84 & 1.78 \\
        & SF600 & 600 &  &  & 4.4 &  &  &  & 2.00 & 1.84 \\        
        \bottomrule
    \end{tabular*}
    }
    \begin{tablenotes}
    \item[*] \footnotesize {$\dagger$ $\alpha$ is taken as the value calibrated from the slump flow test in Section \ref{sec: slump flow test}.}
    \item[*] \footnotesize Note: $t_{\rm{v,\, solid}}$ and $t_{\rm{v,\, fluid}}$ denote the simulated V-funnel flow time using the elasto-viscoplastic solid model and the PR-Bingham fluid model, respectively.
    \end{tablenotes}
\label{table: 4}
\end{table}

It can be observed that $t_v$ increases slightly with rest time. This indicates that, in addition to depending on the plastic viscosity $\eta$, $t_v$ is also influenced by the yield stress through its effect on the apparent viscosity $\eta_a$. Comparing Mix-A and Mix-B shows that Mix-B, which has a higher plastic viscosity, consistently exhibits a larger $t_v$. The PR-Bingham model predicts slightly lower $t_v$ values than the proposed elasto-viscoplastic model, and this difference grows with rest time, reaching a maximum of \hl{0.16 seconds} in the Mix-B-SF600 case. Nevertheless, the overall discrepancy between the two models remains small because, in the flowing regime, the elasto-viscoplastic model produces a velocity profile comparable to the PR-Bingham model, as verified in Section \ref{subsec:poiseuille_flow}.

\begin{figure}[h!]
    \centering
    \includegraphics[width=1\linewidth]{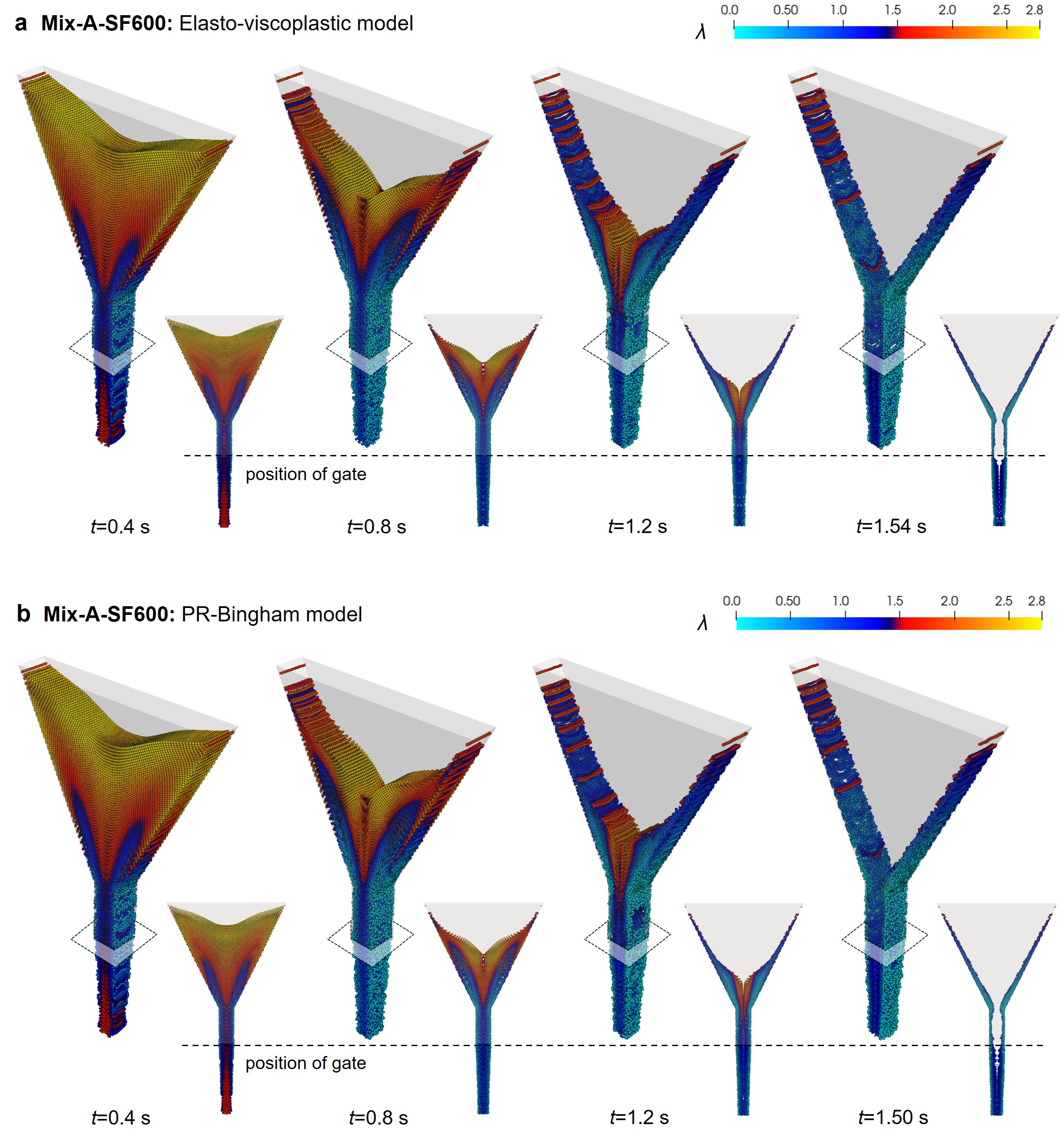}
    \caption{V-funnel test: Simulation results of Mix-A-SF600 using (a) the elasto-viscoplastic model and (b) the PR-Bingham model. Note that the last snapshot shows the results at $T_v$, the time until light can first be seen through the gate.}
    \label{fig: V-funnel-3}
\end{figure}

\begin{figure}[h!]
    \centering
    \includegraphics[width=1\linewidth]{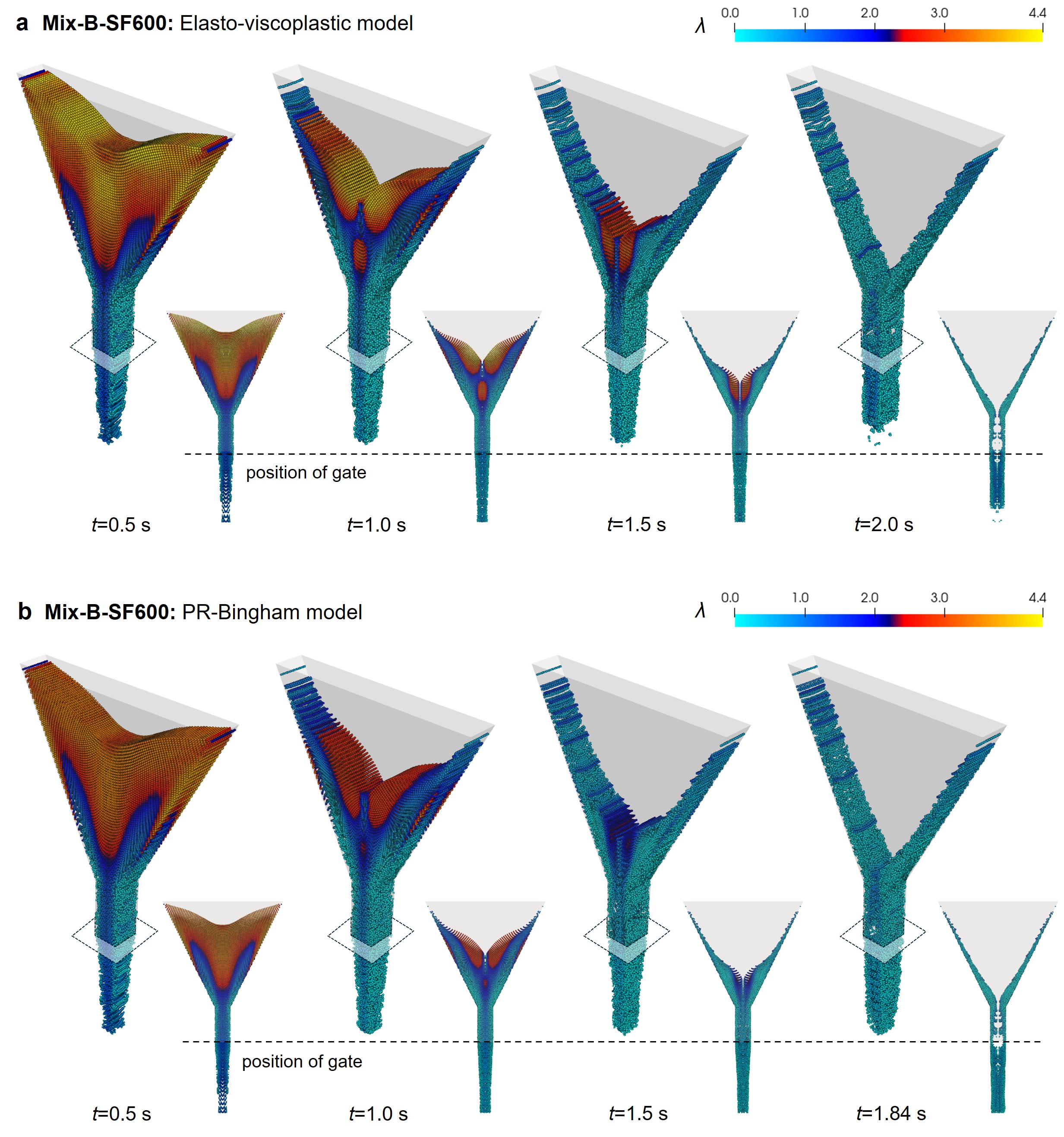}
    \caption{V-funnel test: Simulation results of Mix-B-SF600 using (a) the elasto-viscoplastic model and (b) the PR-Bingham model.}
    \label{fig: V-funnel-4}
\end{figure}

Figs.~\ref{fig: V-funnel-3} and \ref{fig: V-funnel-4} further compare the thixotropy state parameter $\lambda$ from both models and show that their predictions remain close throughout the V-funnel test. These results suggest that, for simulating the flow phase of concrete, the proposed elasto-viscoplastic solid model performs equivalently to the PR-Bingham fluid model. However, unlike the latter, the elasto-viscoplastic model also accurately captures the cessation of flow, providing a more physically realistic representation of the complete flow--rest transition behavior.

\section{Conclusions and outlooks} 
\label{sec: conclusion}

This study developed a novel elasto-viscoplastic thixotropic model for simulating the flow and cessation behavior of fresh concrete. The model was implemented within the Material Point Method (MPM) framework to effectively handle large deformation flows. The key findings of this study can be summarized as follows:
\begin{itemize}
    \item Unlike conventional fluid-based models such as the Bingham model, which rely on ad hoc stopping criteria, the proposed model treats fresh concrete as a rate-dependent elasto-viscoplastic solid within the continuum mechanics framework. This approach naturally captures flow cessation as stresses fall below the yield threshold, enabling a physically realistic transition to a static state.

    \item The integration of Roussel’s thixotropy model allows the simulation of flocculation-induced structural build-up during rest and over time, as well as the breakdown of this structure through shear-induced deflocculation during flow.

    \item An accurate stress integration algorithm based on the return-mapping concept was developed and implemented within the MPM framework, ensuring robust and consistent stress updates across flowing and resting regimes.

    \item Verification tests were performed to assess the model performance under pure flow conditions. In particular, a two-dimensional Poiseuille flow test verified that the proposed elasto-viscoplastic model and the regularized Bingham model yield essentially equivalent results under flowing conditions.

    \item The capabilities of the proposed model were further demonstrated through simulations of standard concrete tests, including slump flow, L-box, and V-funnel tests, with results compared against the PR-Bingham model as well as existing analytical, numerical, and experimental results. The proposed model accurately captures the dynamic flow process, final runout distance, and cessation time. Upon stopping, the plastic shear rate reduces to zero, and the material begins to respond elastically. In contrast, the PR-Bingham fluid model results in perpetual flow due to a non-vanishing shear rate, leading to unphysically overpredicted flow, such as ever-increasing slump diameters in slump flow tests and inaccurate final surface profiles in L-box tests. Furthermore, a single set of thixotropic parameters successfully simulates concrete flow at different flocculation states, demonstrating parameter transferability across different test setups (e.g., from slump flow to L-box to V-funnel tests). The proposed model also captures the reduction of the flocculation state parameter, $\lambda$, during flow and its recovery during rest, as well as localized deflocculation induced by shear near obstacles such as rebars, which is crucial for simulating fresh concrete flow in practical casting scenarios.
\end{itemize}

In summary, the elasto-viscoplastic thixotropic framework proposed in this study provides a physically consistent, numerically robust, and predictive tool for modeling the flow–rest transition in fresh concrete, addressing key limitations of fluid-based approaches. Despite these advancements, the model still has several limitations that suggest valuable directions for future research. First, the adopted Roussel thixotropy model, while practical, remains a simplification; \hl{incorporating more advanced nonlinear and microstructural evolution models could better separate and represent the effects of physical flocculation and chemical hydration}. Second, more accurate boundary and interface modeling is needed to simulate flow around complex geometries and more realistic casting scenarios, such as congested reinforcement, which would benefit from improved boundary-condition imposition and contact algorithms capable of resolving the flow profile near interaction regions \citep{liang2023imposition, liang2024mortar}. Third, the model does not currently account for thermal effects, chemical reactions during resting, or concrete bleeding when water separates from the mixture; extending the framework to include thermo-hydro-chemo-mechanical couplings \citep{baumgarten2019general, yu2024semi, yu2025enhancing, yu2026fully} would enable more comprehensive simulations of concrete behavior during casting and early-age curing. \hl{Last but not least, the present implementation relies on CPU-based parallelization, which may limit its efficiency for long-duration simulations involving extended flocculation histories, such as staged casting or multi-layer 3D printing. Because the constitutive update is evaluated locally at each material point, the framework is naturally suited to GPU parallelization \citep{dong2018large, wang2020massively}, a promising direction for future development.}

\section* {Declaration of competing interest}
\label{Declaration of competing interest}
 The authors declare that they have no known competing financial interests or personal relationships that could have appeared to influence the work reported in this paper.

 \section*{CRediT author contribution statement}
 \textbf{J. Yu:} Methodology, Software, Investigation, Validation, Formal analysis, Visualization, Writing - original draft. \textbf{B. Chandra:} Conceptualization, Methodology, Software, Investigation, Validation, Formal analysis, Visualization, Writing - original draft. \textbf{C. Wilkes}: Resources, Writing - review and editing. \textbf{J. Zhao:} Resources, Supervision, Writing - review and editing. \textbf{K. Soga:} Resources, Supervision, Writing - review and editing.

\section* {Data availability}
The data that support the findings of this study are available from the corresponding author upon reasonable request.

\section* {Acknowledgements}
\label{Acknowledgements} 
The authors thank Dr.~Hao Luo of the University of Cambridge for providing the experimental data presented in Section \ref{sec: Hao's experiment}, and Chris Barker and Duncan Nicholson from Arup London for discussions on concrete modeling. J. Yu acknowledges the support of the Overseas Research Award from the Hong Kong University of Science and Technology for an exchange visit to the University of California, Berkeley. B. Chandra acknowledges Prof.~Ken Kamrin of UC Berkeley for many enlightening discussions on topics related to plasticity theory and modeling.

\footnotesize
\bibliographystyle{elsarticle-harv}
\bibliography{references}

\end{document}